\documentclass[acmlarge]{acmart}
\AtBeginDocument{%
  \providecommand\BibTeX{{%
    \normalfont B\kern-0.5em{\scshape i\kern-0.25em b}\kern-0.8em\TeX}}}

\setcopyright{acmcopyright}
\copyrightyear{2021}
\acmYear{2021}
\acmDOI{10.1145/xxxxxxxxx}

\acmJournal{JOCCH}
\acmVolume{999}
\acmNumber{999}
\acmArticle{999}
\acmMonth{999}

\newcommand{\suzan}[1]{\textcolor{black}{#1}}
\newcommand{\green}[1]{\textcolor{black}{#1}}
\usepackage{siunitx}

\begin{document}

\title{Can BERT Dig It? -- Named Entity Recognition for Information Retrieval in the Archaeology Domain}

\author{Alex Brandsen}
\authornote{Corresponding author}
\email{a.brandsen@arch.leidenuniv.nl}
\orcid{0000-0003-1623-1340}
\affiliation{%
  \institution{Faculty of Archaeology, Leiden University}
  \streetaddress{Einsteinweg 2}
  \city{Leiden}
  \country{The Netherlands}
  \postcode{2333 CC}
}

\author{Suzan Verberne}
\email{s.verberne@liacs.leidenuniv.nl}
\orcid{0000-0002-9609-9505}
\affiliation{%
  \institution{Leiden Institute for Advanced Computer Science, Leiden University}
  \streetaddress{Niels Bohrweg 1}
  \city{Leiden}
  \country{The Netherlands}
  \postcode{2333 CA}
}

\author{Karsten Lambers}
\email{k.lambers@arch.leidenuniv.nl}
\orcid{0000-0001-6432-0925}
\affiliation{%
  \institution{Faculty of Archaeology, Leiden University}
  \streetaddress{Einsteinweg 2}
  \city{Leiden}
  \country{The Netherlands}
  \postcode{2333 CC}
}

\author{Milco Wansleeben}
\email{m.wansleeben@arch.leidenuniv.nl}
\orcid{0000-0001-6432-0925}
\affiliation{%
  \institution{Faculty of Archaeology, Leiden University}
  \streetaddress{Einsteinweg 2}
  \city{Leiden}
  \country{The Netherlands}
  \postcode{2333 CC}
}

\renewcommand{\shortauthors}{Brandsen et al.}

\begin{abstract}

    The amount of archaeological literature is growing rapidly. Until recently, these data were only accessible through metadata search. We implemented a text retrieval engine for a large archaeological text collection ($\sim 658$ Million words). In archaeological IR, domain-specific entities such as locations, time periods, and artefacts, play a central role. This motivated the development of a named entity recognition (NER) model to annotate the full collection with archaeological named entities. 
    In this paper, we present ArcheoBERTje, a BERT model pre-trained on Dutch archaeological texts. We compare the model's quality and output on a Named Entity Recognition task to a generic multilingual model and a generic Dutch model. We also investigate ensemble methods for combining multiple BERT models, and combining the best BERT model with a domain thesaurus using Conditional Random Fields (CRF). 
    We find that ArcheoBERTje outperforms both the multilingual and Dutch model significantly with a smaller standard deviation between runs, reaching an average F1 score of 0.735. The model also outperforms ensemble methods combining the three models. Combining ArcheoBERTje predictions and explicit domain knowledge from the thesaurus did not increase the F1 score. We quantitatively and qualitatively analyse the differences between the vocabulary and output of the BERT models on the full collection and provide some valuable insights in the effect of fine-tuning for specific domains.
    Our results indicate that for a highly specific text domain such as archaeology, further pre-training on domain-specific data increases the model's quality on NER by a much larger margin than shown for other domains in the literature, and that domain-specific pre-training makes the addition of domain knowledge from a thesaurus unnecessary.
    
\end{abstract}



\maketitle

\section{Introduction} 
    \label{sec:intro}

    Like in other domains, archaeologists produce large amounts of text about their research. Besides research leading to scholarly output, commercial archaeology companies survey and excavate areas before developers build there and might destroy the archaeological remains. For each of these investigations, a report is written and stored in a repository. In the Netherlands, more than 4,000 of these documents are produced every year \cite{RCE2017DeErfgoedmonitor}, with the total currently estimated at 70,000. These documents are used to some extent by both academic and commercial archaeologists to do further research.
        
    Currently, this so-called `grey literature' is under-used, as the available search tools only offer metadata search, making searching through these reports time consuming and inaccurate \cite{Habermehl2019OverOnderzoek}. A strong need for better search tools has been well documented in prior work \cite{VandenDries2016IsManagement,Habermehl2019OverOnderzoek,Richards2015TextReports, Brandsen2019}, as the information in the full text of the reports can be of great value. Archaeological information needs are often recall-oriented list questions, consisting of a combination of What, Where and When aspects, e.g. ``Find all cremations from the Early Middle Ages in the Netherlands'' \cite{Brandsen2019}. These are difficult to satisfy as the previously available search interfaces only offer search on the title, a short description, and sometimes information about the dating and type of archaeology encountered (stored in metadata fields), but the latter two are often missing or incorrectly assigned. Archaeologists want to search in more detail, and are often interested in the so called `by-catch': a single find unlike the rest of an excavation. For example, on an excavation yielding mainly Bronze Age material, a single Medieval cremation most likely will not be mentioned in the metadata, making it difficult to retrieve without manually searching through all the PDFs.
    
    To address these needs, we implemented a text retrieval engine for the large collection of archaeological reports in the Netherlands. The retrieval collection contains an export (obtained in 2017) of every PDF file in the DANS repository\footnote{https://easy.dans.knaw.nl/ui/home} with the label `Archaeology'. This totals over 60 thousand documents and 658 Million tokens.
    
    A full text search would alleviate a lot of the current challenges archaeologists face in their search of information, but as \citeauthor{Habermehl2019OverOnderzoek} \cite{Habermehl2019OverOnderzoek} mentions, even in the relatively structured metadata, both synonymy and polysemy are a challenge, which is likely to be even worse in the free text in the body of the documents. 
            
    \begin{itemize} 
        \item Synonymy is a challenge because it leads to a lower recall: as there are numerous ways to write concepts relevant to archaeology, a search for one of these variants will not return the others. Specifically time periods have many synonyms. For example, the `Early Middle Ages' can also be expressed as the `Early Medieval Period', or `Merovingian Period', or as dates that fall within the period, such as `600 CE' and `1400 BP'. 
        \item Polysemy on the other hand, causes precision to be lower because one word can have multiple meanings, causing irrelevant meanings to appear in the search results. A good archaeological example is \textit{Swifterbant}, which is a location, a type of pottery, an excavation event, and a time period. This problem of polysemy causes query ambiguity, as a full-text search engine does not know which meaning the user is looking for in their query, and then also does not know which meaning to retrieve from the corpus.
    \end{itemize}
        
    Automatic query expansion is often used to combat problems with synonymy, either by using thesauri or embeddings to add synonyms and similar terms to a query and increase the recall \cite{Soto2008FuzzyRetrieval,Carpineto2012ARetrieval}. Unfortunately in the case of time periods, this is difficult, as some time periods span thousands or millions of years, and adding each year with multiple variations (AD, BC, CE, BCE, BP) would result in an extremely large query. Polysemy is usually addressed in web search engines by diversifying search results or query suggestions~\cite{Capannini2011EfficientResults,Song2011Post-rankingResults}: for each possible meaning of the ambiguous query, at least one relevant result is shown. For our specific domain, this is not possible because we do not have the large amount of user traffic that generic web search engines have, to be able to learn the different relevant results for any query term.

    Instead, we opt for Named Entity Recognition (NER) to automatically detect archaeological entities in the corpus, and then allow archaeologists to find these using an entity-based query interface, combined with a full text search. The entity search \green{attempts to solve} the polysemy problem, as the user specifies -- in the structured query interface -- which meaning of a word they are looking for, e.g. the Location\footnote{Entity types will be capitalised from here on for clarity.} \textit{Swifterbant}. In this case, \green{only documents where the Location entity \textit{Swifterbant} has been detected will be returned. \suzan{Although this helps the user specifying their query, it also} means that entities that have not been correctly identified will not be returned\suzan{; in other words, errors in the NER output might} propagate to retrieval errors. \suzan{Therefore, to give the user freedom in the query form that best suits their information need, we combine entity search with full-text search. }} 
        
    We have previously published a prototype of our search engine online. The search engine uses ElasticSearch \cite{Gormley2015Elasticsearch:Engine} to index the full text, and in the prototype, entities were automatically labelled with a baseline NER model based on Conditional Random Fields (CRF). The resulting entity-based full-text search was experienced as positive by a focus group of archaeologists \cite{Brandsen2019}. 
        
    However, the baseline NER model offers room for improvement. As prior work on archaeological NER  indicated, CRF with common token-, context- and thesaurus-based features leads to relatively low F1 scores, around 0.50 to 0.60 \cite{Brandsen2020CreatingDomain}. In the last couple of years, transfer learning, and specifically BERT models \cite{Devlin2019BERT:Understanding}, have been used successfully to get state-of-the-art (SotA) results for NER. On general-domain benchmarks the SotA methods yield impressive F1 scores of up to 0.943 \cite{Yamada2020LUKE:Self-attention}. However, in other domains and languages the performance of NER systems is generally lower \cite{Lee2019BioBERT:Mining}.
        
    BERT has not been applied to the archaeology domain yet in any language, and we believe this domain could benefit from context-dependent embeddings due to the above mentioned polysemy. Two generic Dutch BERT models have been released \cite{deVries2019BERTje:Model,Delobelle2020RobBERT:Model} which can help our research. Prior work on language- and domain-specific BERT models reports mixed results on the effect of pre-training on language- and domain-specific data (see Section~\ref{sec:specbert}). In this paper we investigate whether BERT can improve NER in the Dutch archaeology domain, and to what extent further pre-training on domain-specific texts improves the quality of the model. We compare Google's multilingual model \cite{Devlin2019BERT:Understanding}, the Dutch BERTje model \cite{deVries2019BERTje:Model}, and our own ArcheoBERTje model that we further pre-trained on Dutch excavation reports. We do not compare the Dutch RobBERT model as it has a different architecture. We analyse the differences between the three models and we experiment with ensembles to combine multiple models and a domain-specific thesaurus. \green{As there is unfortunately no test collection with relevance assessments available \suzan{for the Dutch archaeology domain}, we do not evaluate the performance of the information retrieval, only the performance of the NER.}
    
    We address the following research question:
        
    \begin{enumerate}
        \item To what extent does further pre-training a BERT model with domain-specific training data improve the model's quality in our highly specific domain?
        \item Can a domain-specific BERT model be improved by adding domain knowledge from a thesaurus in a CRF ensemble model?
        \item What errors are made by the models and what are the differences in predicted entities between the three models?
    \end{enumerate}
        
    The contributions of our paper are three-fold: First, we propose entity-driven full-text search in which the professional user enters a structured query, and documents are filtered for the occurrence of the query entities detected by our new domain-specific BERT model.
    Second, we show that for a highly specific domain such as archaeology, further pre-training on domain-specific data increases the model's quality on NER by a much larger margin than shown for other domains in the literature. 
    Third, we show that the domain-specific BERT model outperforms ensemble methods combining different BERT models, and also outperforms a CRF-based ensemble of BERT with explicit domain knowledge from the archaeological thesaurus. 

    We make our modified training data set, the pre-trained Archeo\-BERTje model, and the fine-tuned ArcheoBERTje model for NER publicly available \cite{Brandsen2021}.\footnote{\url{https://doi.org/10.5281/zenodo.4739063}, also available via the HuggingFace library for ease of use: \url{https://huggingface.co/alexbrandsen}}

\section{Related Work}

    In this section, we first summarise different approaches to NER (knowledge-driven and data-driven), followed by a discussion of related work on NER for document retrieval, on IR and NER in the archaeological domain, and we summarise the prior work on domain-specific BERT models.

    \subsection{Knowledge-driven and Data-driven NER} 

        Early NER systems were knowledge-based, and relied on thesauri and handcrafted rules to detect entities \cite{Rau1991ExtractingText}. These methods are limited by the coverage of the thesaurus. Therefore, data-driven methods have become more popular, typically approaching NER as a supervised machine learning problem.
        
        A highly effective machine learning method is Conditional Random Fields (CRF) \cite{Lafferty2001ConditionalData}, which has become a common baseline for NER. Since 2011, word embeddings have become increasingly important as representations in NER. Especially Word2vec ~\cite{Mikolov2013EfficientSpace} has been used extensively for NER \cite{sienvcnik2015adapting,Seok2016NamedFeature}. These embeddings-based methods typically feed the embeddings to CRF and/or Bi-LSTM algorithms to make NER predictions.
            
        A big shift in NLP was introduced by \citeauthor{Devlin2019BERT:Understanding} \cite{Devlin2019BERT:Understanding}, who presented their BERT (Bidirectional Encoder Representations from Transformers) architecture in 2019. BERT and other contextual embedding architectures are currently achieving SotA results with transfer learning for a large range of NLP tasks, including NER. Two major differences with previous embedding models are (1) that BERT embeddings are contextual, meaning that the same token can have a different embedding based on context, and (2) that it handles out-of-vocabulary words effectively, by dividing tokens into sub-tokens it does have in vocabulary, using the WordPiece \cite{Devlin2019BERT:Understanding} or SentencePiece \cite{Kudo2018SentencePiece:Processing} tokeniser. 
            
        Recent results indicate that ensemble methods that combining generic and domain-specific BERT models \cite{Copara2020NamedModels}, combining BERT with dictionary features \cite{Li2020ChineseMethods}, or adding a CRF on top of BERT \cite{Souza2019PortugueseBERT-CRF} can improve NER quality. In this paper, we investigate whether addition of information from a thesaurus can improve NER in a highly specific domain. 
        
    \subsection{NER for Document Retrieval}    
         
        In the context of document retrieval, NER can play a role in better ranking or filtering documents based on entities in the query. \citeauthor{Guo2009NamedQuery} \cite{Guo2009NamedQuery} were the first to address the task of recognising named entities in queries. They found that, despite queries in web search being short, 70\% of the queries contained a named entity. They classify the entities according to a predefined taxonomy using a weakly supervised topic modelling approach on the query data. \citeauthor{Cowan2015NamedQueries} \cite{Cowan2015NamedQueries} also address NER in queries, but for the travel domain. They use CRF on the queries for extracting the relevant entities.
        
        More recently, the relevance of NER on queries has been emphasised for the e-commerce domain. 
        \citeauthor{Wen2019BuildingSearch} \cite{Wen2019BuildingSearch} and \citeauthor{Cheng2020AnSearch} \cite{Cheng2020AnSearch} both implement end-to-end query analysis methods for e-commerce search; the extracted queries are then used to filter the retrieved products. 
        
        As opposed to the prior work, we do not focus on query analysis but on document analysis; our expert users prefer the use of structured queries, which makes query analysis unnecessary (see Section~\ref{sec:indret}). Our documents, on the other hand, are long and unstructured (as opposed to the products in e-commerce search), making NER on the document side necessary for matching structured queries to the relevant documents.

    \subsection{IR and NER in Archaeology}
        
        As argued by Richards et al. \cite{Richards2015TextReports}, archaeology has great potential for thesaurus-based IR and NER, as it has a relatively well-controlled vocabulary and there are thesauri of archaeological concepts available in multiple languages. However, unlike some other fields, archaeology terminology partly consists of common words, like `pit', `well' and `post'. In addition, words can be archaeological entities or not, depending on the context in which they are used (past or present). For example, the word `road' is not archaeologically relevant in the snippet ``pit next to the main road'', but is part of an archaeological entity in the snippet ``a Roman road from 34 CE''. 
                
        Archaeology has started experimenting with IR relatively recently. The focus of the prior work is on Information or Knowledge Extraction, mainly for automatically generating 
        document metadata. An early study by \citeauthor{Amrani2008AArchaeology} aimed specifically at extracting information for archaeology professionals in a knowledge-based approach \cite{Amrani2008AArchaeology}. A more data-driven approach using machine learning to detect Time Period entities was investigated in the OpenBoek project \cite{Paijmans2010,Paijmans2009WhatDescriptions}, but since then most studies have been knowledge-driven \cite{Jeffrey2009TheContext.,Byrne2010AutomaticText,Vlachidis2013AutomaticLiterature,Vlachidis2017}. 
        
        More recently, \citeauthor{TalboomLeontien2017Itdo} experimented with embeddings in a Bi-LSTM model to recognise zooarchaeological entities (species and specific bones) \cite{TalboomLeontien2017Itdo}. A notable exception to the Information Extraction research we often see in archaeology is the work by \citeauthor{Gibbs2012DigitalAustralia} who created a full-text search engine on a small Australian corpus (roughly 1,000 documents) combined with facets based on manually entered metadata \cite{Gibbs2012DigitalAustralia}. 
                
        So far, NLP in the archaeology domain has not benefitted from BERT-based models. We believe it is a good candidate domain for BERT as the polysemy mentioned in the introduction and the present/past distinction mentioned above should be easier to detect with the context-dependent embeddings that BERT produces.

    \subsection{Language- and Domain-specific BERT Models} \label{sec:specbert}
    
        The original BERT paper \cite{Devlin2019BERT:Understanding} did not only present an English BERT model, but also a multilingual model (multiBERT) trained on data in 104 languages. This model is often used when no single-language model is available \cite{Hakala2019BiomedicalBERT,Moon2019TowardsBERT,Kim2020KoreanBERT}. Research by \citeauthor{Wu2020AreBERT} shows that multiBERT achieved higher accuracy on NER and other NLP tasks than monolingual models trained with comparable amounts of data \cite{Wu2020AreBERT}. \citeauthor{Moon2019TowardsBERT} also showed that fine-tuning multiBERT on a mixed language NER dataset provided better results than fine-tuning on individual languages \cite{Moon2019TowardsBERT}. 
        
        However, recent work has shown that for some languages, multiBERT is outperformed by language-specific BERT models \cite{Nozza2020WhatModels}. For NER, this has been shown for Finnish \cite{Virtanen2019MultilingualFinnish}, Dutch \cite{deVries2019BERTje:Model}, German \cite{Chan2021GermansModel} and Russian \cite{Kuratov2019AdaptationLanguage}, among other languages. 
        
        For specific domains, it has been shown that further pre-training the English BERT-base model on large amounts of text from that domain increases the quality of the model on multiple tasks, although sometimes by a small margin. BioBERT in the biomedical domain shows an increase in F1 for NER of only 0.62\% point \cite{Lee2019BioBERT:Mining}. SciBERT, trained on a large amount of scientific texts from different domains, shows an increase in F1 for NER of 2 to 5\% points, indicating that domain pre-training is useful for NER \cite{Beltagy2020SCIBERT:Text}. They also show that training BERT from scratch with a domain-specific vocabulary does not increase F1 substantially compared to fine-tuning an existing BERT model with an existing generic vocabulary, gaining only 0.6\% points. 
        
        When we look at research done on non-English in a specialised domain like our study, there is little prior work. A study in the Russian cyber-security domain shows that the Russian model (RuBERT) outperformed multiBERT, and further pre-training RuBERT with domain-specific documents yielded the highest F1 \cite{Tikhomirov2020UsingDomain}. In the Spanish biomedical domain, \citeauthor{Akhtyamova2020NamedModel} shows a similar result, although their NER BERT model is trained for 30 epochs, possibly leading to over fitting \cite{Akhtyamova2020NamedModel}.
        
        To our knowledge, we are the first to address domain-specific NER for Dutch, and we are the first to automatically label a large archaeological document collection with our domain-specific BERT model for the purpose of professional search.

\section{Data}
    \label{sec:dataset}
    
    The unlabelled data set we use for further pre-training the Dutch BERTje model to ArcheoBERTje consists of over 60k documents and 658 Million tokens across 16.6 Million sentences, around 2GB of data. The documents mainly consist of survey/excavation reports, but also include other documents such as research plans, appendices, maps and data descriptions. 
    
    The labelled training data we use for NER we created previously \cite{Brandsen2020CreatingDomain}, and consists of fifteen documents that have been annotated by archaeology students, totalling roughly 440k tokens and almost 43k annotated entities across six categories: Artefacts, Time Periods, Locations, Contexts, Materials and Species.\footnote{See \cite{Brandsen2020CreatingDomain} for definitions and examples of these entity types.} The inter-annotator agreement reported is 95\% (average pairwise F1 score), so it is of relatively high quality \cite{Brandsen2020CreatingDomain}. The data is stored in the BIO annotation schema, and is available for download.\footnote{Zenodo repository: \url{http://doi.org/10.5281/zenodo.3544544}}

    \subsection{Pre-processing}
    
        For cross-validation, we divided the fifteen annotated documents across five folds so that each fold has a roughly equal number of tokens. The exact fold split and training data can be found on in the Zenodo repository.
        
        We found that in the data set, sentences often exceed the maximum sequence length of 512 WordPiece tokens. This is not because sentences actually have more than 512 words, but partly because tables and OCRed maps and images create very long `sentences' that are not cut up by the sentence detection algorithm. The other cause is that words that are uncommon outside of archaeology are cut up into many sub-tokens by the WordPiece tokeniser, as they do not exist in the vocabulary (also see Section~\ref{sec:tokenisationissues}).
    
        Since sentences longer than 512 tokens will be trimmed, some of the input tokens will not get a prediction. To counteract this, we wrote a pre-processing script that attempts to break at a punctuation mark (`.', `;' or `,') between the 60th and 90th token and if there are none, it inserts a line break after the 90th token. This shortened the sentences sufficiently to have almost no instances where the sentence was longer than 512 WordPiece tokens.

\section{Methods}

    \subsection{Baseline}
    \label{sec:baseline}
    
        As a baseline, we use the method we published previously \cite{Brandsen2020CreatingDomain}, where we trained a CRF model using common word shape features (e.g. occurrence of uppercase letters, numbers), part-of-speech tags (e.g. noun, verb) and an archaeological thesaurus in a five word window, and performed hyperparameter optimisation. We used the same features, and did 5-fold cross validation, leading to a micro F1 score of 0.62. This is relatively low when comparing the score to NER in other domains, where F1 scores between 0.8 and 0.9 are common \cite{Akhtyamova2020NamedModel,Lee2019BioBERT:Mining}. 
    
    \subsection{Fine-tuning BERT for Dutch Archaeology and NER}
    
        \paragraph{Model training for evaluation} To train ArcheoBERTje, we started with the Dutch BERTje model \cite{deVries2019BERTje:Model} and further pre-trained the model with our complete unlabelled archaeological collection, split into a 90/10 train and validation set.\footnote{We used HuggingFace's \cite{Wolf2020Transformers:Processing} language modelling script version 3.0.2.}  We used the same configuration as BERTje, with a batch size of 4. We decided not to train a model from scratch as previous research showed only minimal increase in quality compared to further pre-training~\cite{Beltagy2020SCIBERT:Text} an existing model, and because our corpus is relatively small and would probably not be enough to train an effective model. 
        
        To fine-tune the BERT models for the NER task, we used the labelled data and 5-fold cross validation as described in Section~\ref{sec:dataset}.\footnote{We used HuggingFace's token classification script version 3.0.2.} For model comparison and to investigate the stability of each model with different random seeds, we trained all three models 10 times per fold, each time using a different seed (1, 2, 4, 8, 16, 32, 64, 128, 254, 512) and report averages over all runs and folds (50 runs in total per BERT model). 
        
        \paragraph{Model for full collection labelling} To create the best possible model for inference on the entire corpus, we performed a grid search across hyperparameters as suggested by \cite{Devlin2019BERT:Understanding}. We optimised the hyperparameters with fold 2 as test set, fold 1 as development set, and the other folds as training set, as this combination had the median F1 score across all models and folds. The grid search yielded the following optimal parameters for our data: 2 training epochs, $5*10^{-5}$ learning rate and 0.1 weight decay. We then fine-tuned the inference model on all labelled data with these hyperparameters. This way we maximise the amount of training data available for training the model that we use to label the full collection.

    \subsection{Ensemble Methods}
   
        As far as we are aware, we are the first to combine a multilingual model, a language-specific model and a domain-specific model into one ensemble method. We evaluate the following ensemble methods (one run
        over 5 folds per ensemble):
        
        \begin{itemize}
            \item Majority voting on the predictions of multiBERT, BERTje and ArcheoBERTje;
            \item CRF that uses the prediction labels of the three models as features;
            \item CRF that uses the prediction labels of ArcheoBERTje only;
            \item CRF that uses the prediction labels of the three models as features, combined with the baseline features;
            \item CRF that uses the prediction labels of ArcheoBERTje only, combined with the baseline features.
        \end{itemize}
        
        \begin{figure*}[t]
            \centering
            \includegraphics[width=1\textwidth]{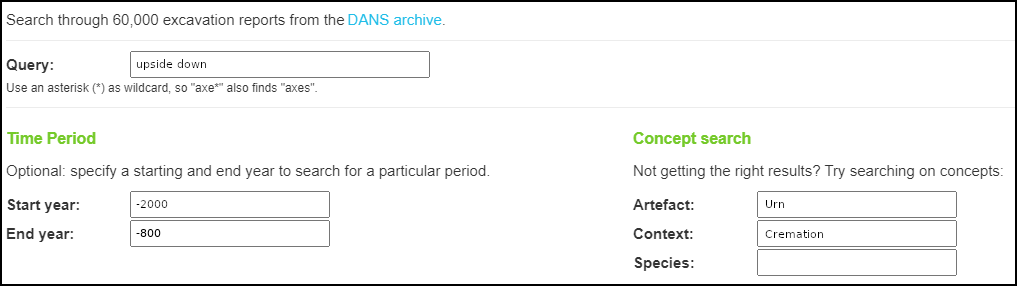}
            \caption{Query interface showing query for ``Artefact: urn AND Context: cremation AND startdate < -2000 AND enddate > -800 AND fulltext: upside down''. Interface and query translated to English for the readers' convenience.}
            \label{fig:interface}
        \end{figure*}
        
        The above mentioned `baseline features' are those adopted from prior work (See Section~\ref{sec:baseline}) and include word shape, part-of-speech tags and thesaurus features. We optimised the hyperparameters of each CRF ensemble with gradient descent using the L-BFGS method, optimising c1 and c2 (the coefficients for L1 and L2 regularisation). The optimisation was run separately for each fold. All CRF ensembles use a 5-token window, taking into account the features from the two tokens before and after the current token. 
        
        The thesaurus we use in our CRF baseline and ensembles is the ABR (\textit{Archeologisch Basisregister}) \cite{Brandt1992ArcheologischBasisregister,Brandsen2020CreatingDomain}, a thesaurus containing time periods (e.g. Bronze Age), artefacts (e.g. axe) and materials (e.g. flint).
        A token is assigned the binary feature `occurs in period/artefact/material list' if it is part of an n-gram that occurs in the thesaurus. So the token `Bronze' would only be assigned a positive value for the feature if the token `age' follows it.

   \subsection{Entity-driven Document Search} 
   \label{sec:indret}
        
        \paragraph{Indexing} 
            
            Before we index the documents, we first run the inference NER model on each page to detect the entities. We then store the entities and full text in a JSON file for each document, together with the relevant metadata  (authors, DOI, coordinates, document type, etc) retrieved from the DANS repository via an API. 
            
            To tackle the synonymy problem for time periods (see Section~\ref{sec:intro}), we use a custom script that translates all extracted Time Period entities to year ranges. It uses regular expressions to convert dates (e.g. `100 BCE', `start of the 9th century') and an extended and customised version of the PeriodO time period gazetteer \cite{Rabinowitz2016MakingProject} to translate Time Periods (e.g. `Bronze Age', `Medieval period'). These date ranges are added to the JSON and can be used to filter results by allowing users to specify a date range in their query. 
            These JSON files are then sent to an instance of ElasticSearch running on a webserver, which indexes them. At the moment, the retrieval unit is a page, so for any query the terms/entities must occur together on a page. We are aware this is not optimal, as search terms might be split across pages. As such, in future work we will index per document section by using a section detection algorithm.
            
        \paragraph{Query Interface and Analysis}

            Our search engine has a faceted search interface in which metadata filters are combined with entity fields and full-text search \cite{Tunkelang2009FacetedSearch}. We have included facets for document type and subject (meta-data fields). In addition, as requested by our target group, we added geographical search via a map functionality, which allows users to draw a rectangle or polygon to search only in a certain region. 
  
            At query time, the user can specify if they are looking for a specific entity type and/or specify a date range in which they are interested. The entities and date range are used to filter the result set, and can be combined with a standard full text search. This allows for relatively complex queries such as ``Artefact: urn AND Context: cremation AND startdate $< -2000$ AND enddate $> -800$ AND fulltext: upside down''. This example is a real request entered by an archaeologist, who was looking for upside down urns in the Bronze Age in or around cremations. Users do not need to use complex query syntax, but can instead define their query by filling in the relevant fields in the graphical user interface, as shown in Figure \ref{fig:interface}.

        \paragraph{Document Ranking} 
            
            Most archaeological information needs are recall-oriented tasks: the users want a complete list and do not mind having irrelevant results in the (top of) the result set \cite{Brandsen2019}. As the focus of our work is on entity-driven search, we opt for the default ElasticSearch ranking model, consisting of TF-IDF and the field-length norm (the shorter the field, the higher the relevance) \cite{ElasticSearch2018TheoryScoring}. The only field included for ranking is the page text content, other fields are only used for filtering.
            
            \suzan{Note that we do not evaluate the ranking, because there is no test collection available yet for Dutch archaeological document retrieval. Therefore, the scope of this paper is limited to the NER and evaluation thereof.}
            
            \begin{table}[b]
            \begin{tabular}{|l|l|l|l|l|}
                \hline
                \textbf{Model} & \textbf{Precision} & \textbf{Recall} & \textbf{F1 (Std.)} & \textbf{Fails} \\ \hline
                CRF Baseline   & \textbf{0.785}               & 0.526                & 0.630 (-)                    & n/a            \\ \hline
                multiBERT      & 0.623               & 0.550                & 0.583 (0.015)           & 4              \\ \hline
                BERTje         & 0.718               & 0.682                & 0.699 (0.005)           & 0              \\ \hline
                ArcheoBERTje   & 0.743      & \textbf{0.729}       & \textbf{0.735 (0.004)}  & 0              \\ \hline
            \end{tabular}
            \caption{Micro average precision, recall and F1 score at token level (B and I labels), over 10 runs with different seeds, for each of the 5 folds (50 runs total). Standard deviation of F1 over the 10 runs is added in brackets. Standard deviation of precision and recall lies between 0.006 and 0.020. CRF has no standard deviation as it is deterministic. The `Fails' column indicates the number times the model failed to learn (F1 = 0).}
            \label{tab:results}
        \end{table}

\section{Results}

    \begin{table*}[t]
        \begin{tabular}{|l|r|r|r|}
        \hline
        Ensemble                                                     & \textbf{Precision} & \textbf{Recall} & \textbf{F1} \\ \hline
        ArcheoBERTje solo                                            & 0.743                   & \textbf{0.729}       & 0.735            \\ \hline
        Majority Voting                                              & 0.784                   & 0.695                & 0.737            \\ \hline
        CRF with 3 BERT model prediction labels as features          & 0.786                   & 0.683                & 0.731            \\ \hline
        CRF with only ArcheoBERTje predictions as features           & 0.786                   & 0.717                & \textbf{0.750}   \\ \hline
        CRF with 3 BERT model prediction labels + baseline  features & \textbf{0.795}          & 0.644                & 0.712            \\ \hline
        CRF with ArcheoBERTje prediction labels + baseline  features & 0.793                   & 0.649                & 0.714            \\ \hline
        \end{tabular}
        \caption{Micro F1 score, precision and recall for the five ensemble methods, for one run over five folds. ArcheoBERTje results (ten runs, five folds) added for comparison. 
        The baseline features are the word- and context-based features used for CRF in prior work.}
        \label{tab:ensemblef1s}
    \end{table*}

    \subsection{Model Stability and Quality}
    \label{sec:stability}
    
        Table \ref{tab:results} shows the micro average precision, recall and F1 score for the three BERT models, compared to the CRF baseline. We also show the average standard deviation over 10 runs with different seeds for 5 folds. The standard deviation between runs is very low, between 0.015 and 0.004. The recent work by \citeauthor{Tikhomirov2020UsingDomain} reports a standard deviation of 0.015 to 0.008, similar to our results \cite{Tikhomirov2020UsingDomain}. When comparing the predicted labels of each of the models in a pairwise manner, the differences are significant according to McNemar's test ($\chi^2$ between 650 and 4276, $p <0.00001$).
    
\begin{figure}[b]
            \centering
            \includegraphics[width=0.48\textwidth]{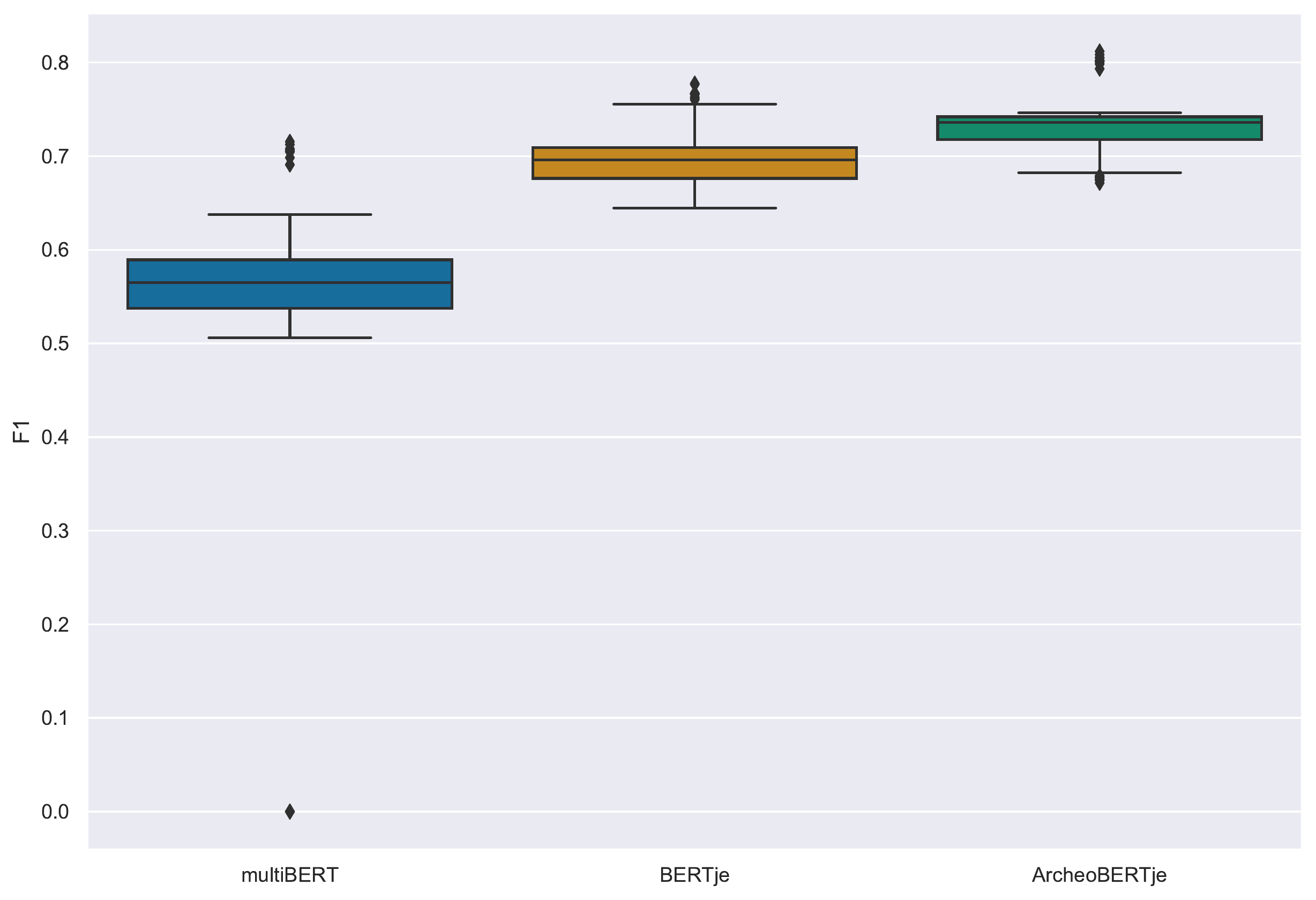}
            \caption{Distribution of F1 scores over ten runs with different seeds, for each of the 5 folds (50 runs per model). The zero scores for multiBERT are runs where the model failed to learn. }
            \label{fig:distribution-boxplot}
        \end{figure}

        Figure \ref{fig:distribution-boxplot} shows the distribution of F1 scores over the 50 runs per model in a boxplot. Here we again see that the standard deviation is low, and that ArcheoBERTje consistently outperforms the other two models. The F1 scores of 0 for multiBERT are outliers, and we assume these are caused by the ADAM optimiser getting stuck in a local minimum where the loss does not decrease. In this local minimum, predicting the majority class (O) seems to yield the highest accuracy, but of course O labels are not taken into account when calculating an F1 score for NER, so we get a score of zero. This can be solved by changing the learning rate, but this would not change the overall view that BERTje and ArcheoBERTje outperform multiBERT, so we did not investigate further on fixing this for multiBERT.
        
        The low standard deviations for ArcheoBERTje indicate that further pre-training with domain-specific data does not only increase the model quality on average, but also makes the model more stable, reducing the chance of getting a sub-optimal model in a run.
        
        Another way to compare the models is by looking at differences between the errors made. In Table \ref{tab:errorfreqs} we report the top 10 most frequent error combinations for the three models. Here we can see that quite often, BERTje and ArcheoBERTje have similar predictions (whether correct or not), while multiBERT predicted a different label. We see that multiBERT often misses Locations (LOC), Artefacts (ART) and Species (SPE), and sometimes predicts entities that are not there. The first error combination where ArcheoBERTje outperforms BERTje is number 9, having correctly predicted B-ARTs while the other 2 models do not.  In Sections~\ref{sec:anafull} and~\ref{sec:errorana} we further analyse the output and errors made by the ArcheoBERTje model to provide insight in the model's behaviour.
    
        \begin{table}[b]
        \begin{tabular}{|l|l|l|l|l|}
        \hline
        \textbf{Freq.} & \textbf{True} & \textbf{multiBERT}           & \textbf{BERTje}              & \textbf{ArcheoBERTje}        \\ \hline
        1137               & B-LOC               & {\color[HTML]{FE0000} O}     & B-LOC                        & B-LOC                        \\ \hline
        1122               & B-ART               & {\color[HTML]{FE0000} O}     & B-ART                        & B-ART                        \\ \hline
        1015               & O                   & {\color[HTML]{FE0000} B-ART} & O                            & O                            \\ \hline
        575                & B-SPE               & {\color[HTML]{FE0000} O}     & B-SPE                        & B-SPE                        \\ \hline
        561                & O                   & {\color[HTML]{FE0000} B-LOC} & O                            & O                            \\ \hline
        466                & B-PER               & {\color[HTML]{FE0000} O}     & B-PER                        & B-PER                        \\ \hline
        429                & O                   & O                            & {\color[HTML]{FE0000} B-ART} & {\color[HTML]{FE0000} B-ART} \\ \hline
        425                & I-PER               & {\color[HTML]{FE0000} O}     & I-PER                        & I-PER                        \\ \hline
        402                & B-ART               & {\color[HTML]{FE0000} O}     & {\color[HTML]{FE0000} O}                            & {\color[HTML]{000000} B-ART} \\ \hline
        373                & O               & {\color[HTML]{FE0000} I-PER}     & O                            & O \\ \hline
        \end{tabular}
        \caption{The 10 most frequent error combinations between the 3 models for which at least one model has the correct prediction. Errors are marked in red.}
        \label{tab:errorfreqs}
        \end{table}

    \subsection{Ensembles}
        \label{sec:ensembles}
        
         \begin{table*}[t]
            \begin{tabular}{|l|r|r|l|}
            \hline
            \textbf{Entity} & \textbf{Total} & \textbf{Unique} & \textbf{Top 5}                                           \\ \hline
            Artefacts       & 2,520,492      & 53,675          & pottery, charcoal, flint, bone, brick                    \\ \hline
            Contexts        & 1,602,124      & 21,319          & pit, ditch, posthole, well, house                        \\ \hline
            Materials       & 457,031        & 6,146           & wooden, flint, wood, metal, bronze                       \\ \hline
            Locations & 3,488,698      & 147,077         & nederland, ' , groningen, noord - brabant, gelderland    \\ \hline
            Species         & 928,437        & 34,540          & cow, hazel, sheep, goat, pig                             \\ \hline
            Time Periods    & 4,698,323      & 98,445          & roman period, iron age, 150 - 210, late medieval, modern \\ \hline
            Total           & 13,695,105     & 361,202         &                                                          \\ \hline
            \end{tabular}
            \caption{Overview of entities detected in the entire corpus, showing total and unique counts, plus the top 5 for each entity (translated from Dutch where relevant).}
            \label{tab:entities}
        \end{table*}
        
        Table~\ref{tab:ensemblef1s} shows the results of the ensemble methods.\footnote{As the standard deviation between multiple runs is low, combining multiple runs of the same model in an ensemble model is very unlikely to increase the F1 score, at the expense of a vastly increased computing time and cost. Hence we do not apply this approach.} The highest F1 (0.750) is obtained by a CRF model that only uses the ArcheoBERTje predictions as features. This is  higher than the F1 we reported for ArcheoBERTje without CRF (0.735, Table~\ref{tab:results}). This increase in F1 is caused by a higher precision, but at the cost of a lower recall than ArcheoBERTje alone.
        
        The highest precision is obtained by the CRF ensemble with the baseline features combined with the predicted labels from all three models. The highest recall is achieved by ArcheoBERTje solo.\footnote{For general domain Portuguese NER, \citeauthor{Souza2019PortugueseBERT-CRF} show the same pattern: Portuguese BERT has the highest recall, while combining BERT with CRF yields the highest precision and F1 \cite{Souza2019PortugueseBERT-CRF}.} Given the recall-oriented nature of professional search tasks like ours, we prioritise recall over precision for the NER labelling, and use ArcheoBERTje for labelling the full collection.

    \subsection{Analysis of the Retrieval Collection} 
        \label{sec:anafull}
        
        After labelling the full retrieval collection with ArcheoBERTje, we analyse the extracted entities. Table \ref{tab:entities} shows for each entity type the total frequency and the amount of unique entities. We also show the top 5 entities extracted for each type (translated from Dutch to English). 
        
        As we already mentioned in the introduction, archaeologists are interested in the What, Where and When of excavations. And so we see that Artefacts, Locations and Time Periods are the most common entities. 
        
        \begin{itemize}
            \item For Artefacts, we see that pottery and flint are common, which we expected, but apparently also charcoal, which we did not expect, but could be explained by the use of carbon dating, which often uses charcoal as a sample.
            \item In the Locations category, we see that the second most common entity is an apostrophe ('). While this is obviously not a location, luckily it will not affect retrieval as it is not something users would search for, 
            and ElasticSearch does not include apostrophes in its index, so it would not match any documents. We speculate that ArcheoBERTje mislabels apostrophes as locations because of the occurrence of apostrophes in some Dutch place names (e.g. \textit{'s Hertogenbosch}). 
            \item For Time Periods, the only unexpected entry in the top 5 is ``150 - 210''. When we investigated this further, we found this is actually a soil grain size used in coring reports, which have been incorrectly labelled as a time period by ArcheoBERTje. 150-210 µm is the grain size for medium course sand, apparently the most common grain size in the Netherlands. When we look further down the Time Period top 100, we also see other common grain sizes: 210-300, 105-150 and 105-210. This is an issue when searching for archaeology between 105 and 300 CE, as these irrelevant coring reports will also be returned. We believe that these errors are made because these numbers come from tables, and as such do not have any sentence context, making them difficult to predict correctly. The most likely way to fix this is by making a post-processing correction on the extracted entities. This is something we will improve in the next version of our NER method. 
        \end{itemize}
        
        \begin{figure}[b]
            \centering
            \includegraphics[width=1\textwidth]{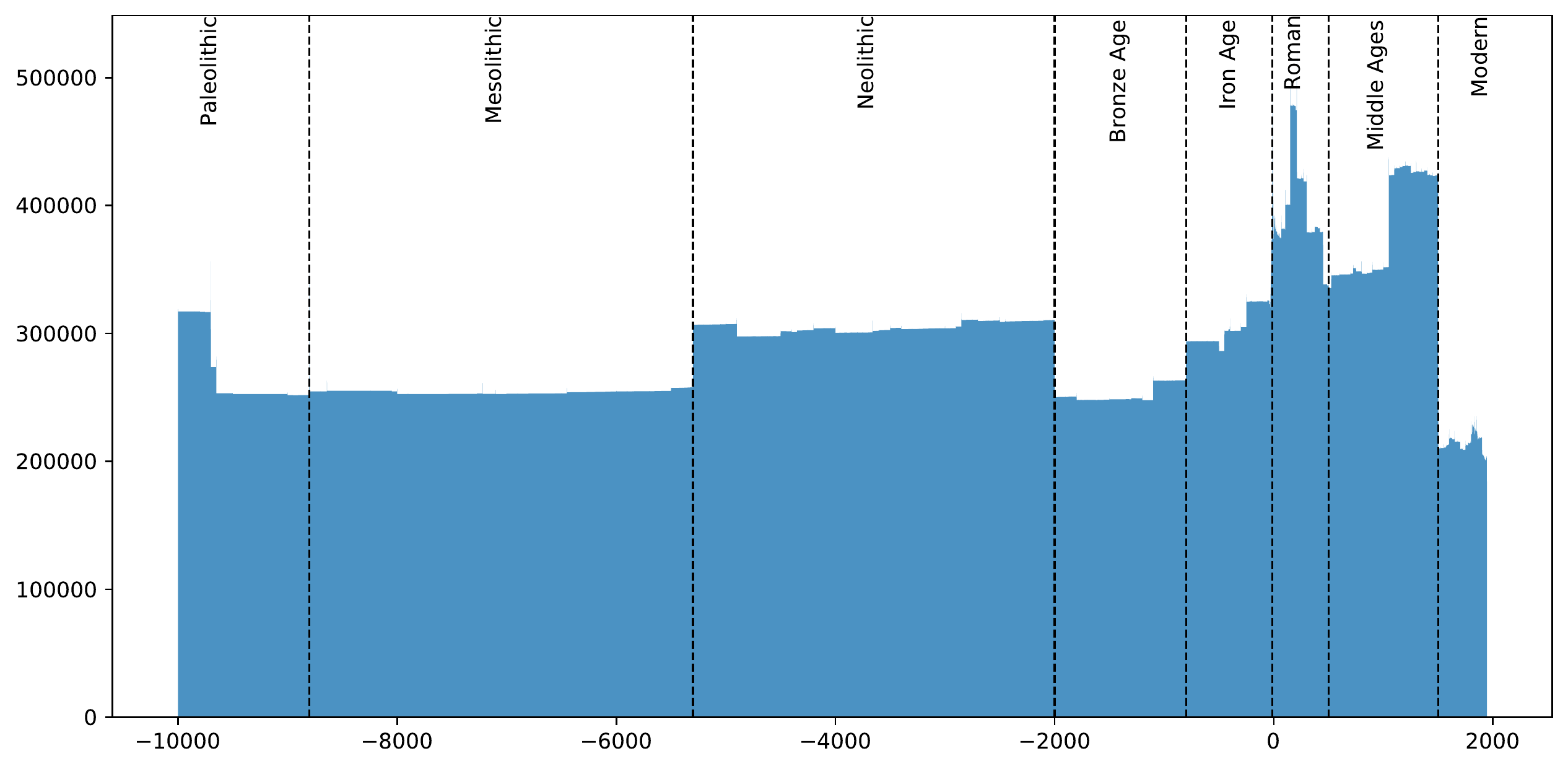}
            \caption{
                Graph showing for each year in each detected time period, how often it occurs in our data set, labelled by ArcheoBERTje. For clarity, years before 10,000 BCE are not included. Major time periods are denoted with dashed lines.
            }
            \label{fig:years}
        \end{figure}
        
        The grain sizes are also clearly visible in Figure \ref{fig:years}, in which we have plotted the frequency of years found in entities in the corpus. The figure shows a number of plateaus, indicating the use of time periods instead of single dates, i.e. the last plateau is the Late Middle Ages ending in 1500 CE. These plateaus are not completely flat as single dates and sub-periods can cause spikes and smaller sub-plateaus.

        The thin spike just after the year 0 can probably be attributed to misclassified entities, i.e. the `10' in `10-02-2006' being labelled by ArcheoBERTje as a Time Period and translated to 10 CE. Other than this we see a big plateau in the middle (5300--2000 BCE), which represents the Neolithic. This indicates that a large amount of data is available describing this period in the Stone Age.
        
        \begin{figure}[t]
            \centering
            \includegraphics[width=0.48\textwidth]{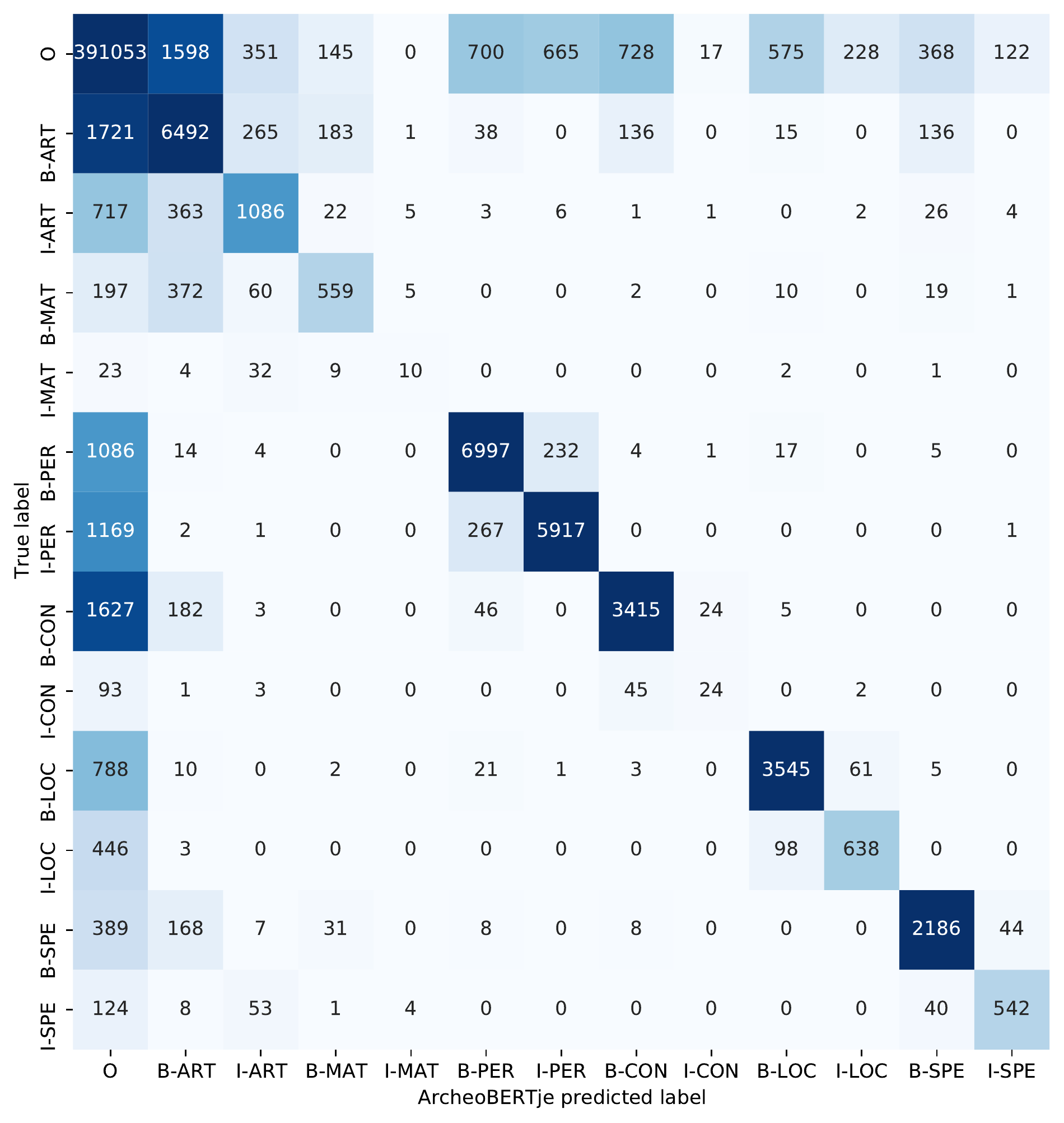}
            \caption{Confusion matrix between true labels and ArcheoBERTje predictions.}
            \label{fig:confusion}
        \end{figure}

\section{Discussion}

    \subsection{Error Analysis} 
        \label{sec:errorana}
        
        Figure \ref{fig:confusion} shows the confusion matrix between labels predicted by ArcheoBERTje and the true labels. The diagonal line, and the first row and column are typical for NER. The diagonal shows the true positives, the top row is where the model predicted an entity where there isn't one, and the first column is where the model predicted O where there should be an entity. We also see the I / B label confusion quite clearly, mainly for Time Periods and Locations, where the model predicts an I instead of a B, or the other way around.
        
        A more interesting error is the confusion between Materials and Artefacts.
        This is caused by words like ``flint'', which can be both an Artefact (``a piece of flint'') or a Material (``a flint axe''). In Dutch, ``pottery'' has the same issue. Even archaeologists struggle with distinguishing between the two \cite{Brandsen2020CreatingDomain}, so it is unsurprising ArcheoBERTje finds this difficult as well.
        
        Table \ref{tab:f1-per-entity} shows the evaluation per entity type. In general, the I labels are more difficult to predict, and Materials are more difficult than the other entities. In fact, Materials are currently not included in the search engine, as archaeologists find it difficult to differentiate between Materials and Artefacts in their queries, so this will not affect retrieval quality. When we remove Materials from the \green{overall micro} F1 score calculation, we get an increase of only around 0.01, as there are only a small number in our training data, around 3000. 
        
        \begin{table}[t]
            \begin{tabular}{|l|r|r|r|}
            \hline
                           & \textbf{Precision} & \textbf{Recall} & \textbf{F1} \\ \hline
            \textbf{B-ART (Artefacts)} & 0.704              & 0.722           & 0.713       \\ \hline
            \textbf{I-ART } & 0.582              & 0.486           & 0.530       \\ \hline
            \textbf{B-CON (Contexts)} & 0.787              & 0.644           & 0.708       \\ \hline
            \textbf{I-CON } & 0.358              & 0.143           & 0.204       \\ \hline
            \textbf{B-MAT (Materials)} & 0.587              & 0.456           & 0.514       \\ \hline
            \textbf{I-MAT} & 0.400              & 0.123           & 0.189       \\ \hline
            \textbf{B-LOC (Locations)} & 0.831              & 0.799           & 0.815       \\ \hline
            \textbf{I-LOC} & 0.685              & 0.538           & 0.603       \\ \hline
            \textbf{B-SPE (Species)} & 0.785              & 0.769           & 0.777       \\ \hline
            \textbf{I-SPE } & 0.759              & 0.702           & 0.729       \\ \hline
            \textbf{B-PER (Time Periods)} & 0.866              & 0.837           & 0.851       \\ \hline
            \textbf{I-PER} & 0.867              & 0.804           & 0.835       \\ \hline
            \textbf{Macro Average} & 0.684              & 0.585           & 0.622       \\ \hline
            \textbf{Micro Average} & 0.784              & 0.731           & 0.757       \\ \hline
            \end{tabular}
            \caption{ArcheoBERTje precision, recall and F1 score for each label.}
            \label{tab:f1-per-entity}
        \end{table}
        
        When we look at some of the errors made by ArcheoBERTje in more depth, we find some interesting patterns. For example, for missing B-ART labels, many errors are adjectives that were assigned the O label, e.g. for ``big axe'' or ``complete pot'', the adjectives are labelled O, and axe / pot are labelled B-ART. This error is not surprising as most archaeologists would probably find it difficult to define these entities as well. In addition, users are  more likely to only search for the base artefact and not include an adjective, so they would search for ``pot'' not ``complete pot''. In a pilot study evaluating our archaeological search engine, we analysed users' search behaviour, and found that of the 148 issued queries, none included an adjective.\footnote{Extension and publication of this user study is part of our future work.}
        
        For Time Periods, we again see that adjectives are missed from the start of an entity, but also prepositions. Some examples include ``from'', ``between'' and ``start of''. Also we find that connecting words between Time Periods are missed, such as ``and'', ``or'' and ``±'' (used to denote the standard deviation of a carbon dating). While this does cause some noise, missing adjectives/prepositions or connecting words are not a huge issue if the main period has been detected. I.e. for ``start of 10th century'', if we miss ``start of'' this means the year range is 900 to 1000 CE, instead of 900 to 925 CE. Again, as archaeologists care more about recall than precision, this should not hinder their search.
        
        The predicted Context\footnote{For clarity, Contexts are defined as an anthropogenic structures or objects that can contain Artefacts, i.e. rubbish pits, burials, houses, and so on.} entities also have some interesting anomalies. In particular, we analysed the top 10 most misclassified tokens and we found that these are all words that can denote contemporary objects (and thus not a Context) or actual (pre-)historical Contexts. An example is ``\textit{put}'', which can mean a trench dug by archaeologists, or a water well found in an excavation. Other examples are ``house'', ``church'', ``ditch'', ``mine'' and ``settlement''. It seems that even with the context-dependent embeddings BERT produces, these ambiguous words are still a challenge. 
        
        A special case is the word ``\textit{poel}'' (pond). We see that this token is always labelled as O while it is in fact a Context. When we checked the sentences this word occurs in, we see they are all very typical of Contexts, i.e. ``we found pottery in the pond'', which is similar to sentence structures of other Contexts that are classified correctly. The only possible explanation we can find is that the word \textit{poel} only occurs in one of the documents, so when this document is in the test set, the word does not occur at all in the train or dev set. This confirms the importance of creating train-test splits on the document level, to avoid overfitting.
        
        More generally speaking, we see that the BERT models make impossible B and I predictions, i.e. an I label without a B label for the previous token. Unlike CRF, which learns the probabilities of two labels occurring after one another, BERT sees every token as an individual classification task without taking into account the predicted label of the previous token. This might explain why the CRF model with ArcheoBERTje labels as features (see Table \ref{tab:ensemblef1s}) outperforms ArcheoBERTje on precision, as it corrects some of these mistakes.

    \subsection{Tokenisation Issues}
        \label{sec:tokenisationissues}
    
        The vocabulary of a BERT model is determined by the collection used for pre-training. The WordPiece tokeniser optimises the set of (sub-word) tokens to maximise the coverage of the collection's vocabulary. The same tokenisation is applied to the input sentences at inference. An example is shown below, where we compare tokenisation with the multiBERT and BERTje vocabularies. We see that target entities (``\textit{Swifterbant}'', ``\textit{aardewerkscherven}'' and ``\textit{Midden Neolithicum}'') are split up into three or more sub-tokens by the multiBERT and BERTje tokenisers. 
        
        \vspace{6pt}
        \noindent \textbf{Original sentence:} 
        
        \noindent ``\emph{In put twee werden 3 Swifterbant aardewerkscherven aangetroffen uit het Midden Neolithicum.}'' (``In trench two, 3 Swifterbant pottery shards from the Middle Neolithic were found.'')
        
        \vspace{6pt}
        \noindent \textbf{multiBERT tokenisation (23 tokens): }
        
        \noindent In put twee werden 3 Swift \#\#er \#\#bant aarde \#\#werks \#\#cher \#\#ven aan \#\#get \#\#roffen uit het Midden Neo \#\#lit \#\#hic \#\#um .
        
        \vspace{6pt}
        \noindent \textbf{BERTje tokenisation (20 tokens), also used for ArcheoBERTje:}

        \noindent In put twee werden 3 Swift \#\#er \#\#ban \#\#t aardewerk \#\#scher \#\#ven aangetroffen uit het Midden Neo \#\#lith \#\#icum .
        
        \vspace{6pt}
        \noindent As additional analysis, we trained a SentencePiece tokeniser on our archaeological collection, with the same vocabulary size as the BERTje model (30k). 
        \vspace{6pt}
        
        \noindent \textbf{Archaeology tokenisation (14 tokens):}
        
        \noindent In put twee werden 3 Swifterbant aardewerk \#\#scherven aangetroffen uit het Midden Neolithicum .

        \vspace{6pt}
        The examples show that a more specific pre-training corpus would lead to more complete domain words. However, our collection is small for such from-scratch pre-training and the experiments in the sciBERT paper have shown that even a much larger pre-training collection only gives a +0.6\% point F1 increase compared to further pre-training the generic model~\cite{Beltagy2020SCIBERT:Text}.         
        
        Understandably, the problem of input sequences longer than 512 tokens was occurring more often with the multilingual model, as the vocabulary (with fixed size) is not solely Dutch. This means that many less common Dutch words are not in the vocabulary, and are cut into many sub-tokens by the WordPiece tokeniser. This effect is aggravated by the Dutch language having a lot of compound words and a much longer average word length (4.8 in English \cite{Norvig2013EnglishRevisited} vs. 8 in Dutch \cite{Corstius1981OpperlandseLetterkunde}).

\section{Conclusion}

    In this paper, we have evaluated BERT models for Named Entity Recognition in the Dutch archaeological domain\suzan{, with the purpose of improving our archaeological search engine}. We implemented the search engine for a large archaeological text collection, with a structured query interface that allows the specification of entity types. The document collection is automatically annotated with archaeological named entities such as Location, Time Period, and Artefact. 
    
    In response to our research questions, first, we found that fine-tuning a BERT model with domain-specific training data improves the model's quality by a large margin for the archaeological domain, larger than in related work addressing domain-specific BERT models. We achieve an average F1 of 0.735 after hyper-parameter optimisation, and very small standard deviations over runs with different random seeds.
    
    Second, the domain-specific BERT model was superior in F1 and recall than an ensemble combining multiple BERT models, and could not be further improved by adding domain knowledge from a thesaurus in a CRF ensemble model. This indicates that after pre-training and fine-tuning on a domain-specific collection, the BERT model already covers the relevant information from the domain thesaurus. We did find a higher \emph{precision} when we combined all three BERT models in a CRF model and added domain knowledge. However, as almost all information needs in archaeology are recall-oriented, and combining models is computationally expensive and environmentally taxing \cite{Strubell2020EnergyNLP}, we opt for the ArcheoBERTje model for labelling the full retrieval collection.
    
    Third, our error analysis shows that there is confusion between the Artefact and Material entities, similar to what humans experienced in the annotation process. For Artefacts and Time Periods, a common error is missing the adjective or preposition in an entity. The detection of Time Periods is a bit noisy, with other non-year numbers erroneously labelled as time ranges.  Context entities such as ``house'' and ``ditch'' are difficult for the models to distinguish from non-entity words. Creating train-test splits on the document level is important to avoid overfitting, as the consistently misclassified Context ``\textit{poel}'' shows, which only occurs in one document. An analysis of tokenisation by each of the models indicates that the multiBERT model is hampered by the rough tokenisation, splitting many relevant terms in sub-words. 
    
    In the near future, we will evaluate the entity-driven search engine with users, both in a controlled experiment and in natural search contexts. We will also investigate entity-based query suggestion. Once entities are mapped to a thesaurus or embedded in a semantic space, this allows for query improvement by suggesting parent or sibling entities in the thesaurus or nearest-neighbours in the embedding space.

\bibliographystyle{ACM-Reference-Format}
\bibliography{final.bib}


\begin{thebibliography}{58}


\ifx \showCODEN    \undefined \def \showCODEN     #1{\unskip}     \fi
\ifx \showDOI      \undefined \def \showDOI       #1{#1}\fi
\ifx \showISBNx    \undefined \def \showISBNx     #1{\unskip}     \fi
\ifx \showISBNxiii \undefined \def \showISBNxiii  #1{\unskip}     \fi
\ifx \showISSN     \undefined \def \showISSN      #1{\unskip}     \fi
\ifx \showLCCN     \undefined \def \showLCCN      #1{\unskip}     \fi
\ifx \shownote     \undefined \def \shownote      #1{#1}          \fi
\ifx \showarticletitle \undefined \def \showarticletitle #1{#1}   \fi
\ifx \showURL      \undefined \def \showURL       {\relax}        \fi
\providecommand\bibfield[2]{#2}
\providecommand\bibinfo[2]{#2}
\providecommand\natexlab[1]{#1}
\providecommand\showeprint[2][]{arXiv:#2}

\bibitem[\protect\citeauthoryear{Akhtyamova}{Akhtyamova}{2020}]%
        {Akhtyamova2020NamedModel}
\bibfield{author}{\bibinfo{person}{Liliya Akhtyamova}.}
  \bibinfo{year}{2020}\natexlab{}.
\newblock \showarticletitle{{Named Entity Recognition in Spanish Biomedical
  Literature: Short Review and Bert Model}}. In \bibinfo{booktitle}{\emph{26th
  Conference of Open Innovations Association (FRUCT)}}.
  \bibinfo{publisher}{IEEE Computer Society}, \bibinfo{address}{Yaroslavl,
  Russia}, \bibinfo{pages}{1--7}.
\newblock
\showISBNx{9789526924427}
\showISSN{23057254}
\urldef\tempurl%
\url{https://doi.org/10.23919/FRUCT48808.2020.9087359}
\showDOI{\tempurl}


\bibitem[\protect\citeauthoryear{Amrani, Abajian, and Kodratoff}{Amrani
  et~al\mbox{.}}{2008}]%
        {Amrani2008AArchaeology}
\bibfield{author}{\bibinfo{person}{A Amrani}, \bibinfo{person}{V Abajian},
  {and} \bibinfo{person}{Y Kodratoff}.} \bibinfo{year}{2008}\natexlab{}.
\newblock \showarticletitle{{A chain of text-mining to extract information in
  archaeology}}. In \bibinfo{booktitle}{\emph{Information and Communication
  Technologies: From Theory to Applications, ICTTA 2008.}}
  \bibinfo{address}{Damascus, Syria}, \bibinfo{pages}{1--5}.
\newblock
\urldef\tempurl%
\url{https://doi.org/10.1109/ICTTA.2008.4529905}
\showDOI{\tempurl}


\bibitem[\protect\citeauthoryear{Beltagy, Lo, and Cohan}{Beltagy
  et~al\mbox{.}}{2020}]%
        {Beltagy2020SCIBERT:Text}
\bibfield{author}{\bibinfo{person}{Iz Beltagy}, \bibinfo{person}{Kyle Lo},
  {and} \bibinfo{person}{Arman Cohan}.} \bibinfo{year}{2020}\natexlab{}.
\newblock \showarticletitle{{SCIBERT: A pretrained language model for
  scientific text}}. In \bibinfo{booktitle}{\emph{EMNLP-IJCNLP 2019 - 2019
  Conference on Empirical Methods in Natural Language Processing and 9th
  International Joint Conference on Natural Language Processing, Proceedings of
  the Conference}}. \bibinfo{publisher}{Association for Computational
  Linguistics}, \bibinfo{address}{Hong Kong, China}.
\newblock
\showISBNx{9781950737901}
\urldef\tempurl%
\url{https://doi.org/10.18653/v1/d19-1371}
\showDOI{\tempurl}


\bibitem[\protect\citeauthoryear{Brandsen}{Brandsen}{2021}]%
        {Brandsen2021}
\bibfield{author}{\bibinfo{person}{A Brandsen}.}
  \bibinfo{year}{2021}\natexlab{}.
\newblock \showarticletitle{{ArcheoBERTje - A Dutch BERT model for the
  Archaeology domain}}.
\newblock \bibinfo{journal}{\emph{Zenodo Repository}} (\bibinfo{year}{2021}).
\newblock
\urldef\tempurl%
\url{https://doi.org/10.5281/zenodo.4739063}
\showDOI{\tempurl}


\bibitem[\protect\citeauthoryear{Brandsen, Lambers, Verberne, and
  Wansleeben}{Brandsen et~al\mbox{.}}{2019}]%
        {Brandsen2019}
\bibfield{author}{\bibinfo{person}{Alex Brandsen}, \bibinfo{person}{Karsten
  Lambers}, \bibinfo{person}{Suzan Verberne}, {and} \bibinfo{person}{Milco
  Wansleeben}.} \bibinfo{year}{2019}\natexlab{}.
\newblock \showarticletitle{{User Requirement Solicitation for an Information
  Retrieval System Applied to Dutch Grey Literature in the Archaeology
  Domain}}.
\newblock \bibinfo{journal}{\emph{Journal of Computer Applications in
  Archaeology}} \bibinfo{volume}{2}, \bibinfo{number}{1} (\bibinfo{date}{3}
  \bibinfo{year}{2019}), \bibinfo{pages}{21--30}.
\newblock
\urldef\tempurl%
\url{https://doi.org/10.5334/jcaa.33}
\showDOI{\tempurl}


\bibitem[\protect\citeauthoryear{Brandsen, Verberne, Wansleeben, and
  Lambers}{Brandsen et~al\mbox{.}}{2020}]%
        {Brandsen2020CreatingDomain}
\bibfield{author}{\bibinfo{person}{Alex Brandsen}, \bibinfo{person}{Suzan
  Verberne}, \bibinfo{person}{Milco Wansleeben}, {and} \bibinfo{person}{Karsten
  Lambers}.} \bibinfo{year}{2020}\natexlab{}.
\newblock \showarticletitle{{Creating a Dataset for Named Entity Recognition in
  the Archaeology Domain}}. In \bibinfo{booktitle}{\emph{Proceedings of The
  12th Language Resources and Evaluation Conference}}.
  \bibinfo{publisher}{European Language Resources Association},
  \bibinfo{address}{Marseille, France}, \bibinfo{pages}{4573–4577}.
\newblock


\bibitem[\protect\citeauthoryear{Brandt, Drenth, Montforts, Proos, Roorda, and
  Wiemer}{Brandt et~al\mbox{.}}{1992}]%
        {Brandt1992ArcheologischBasisregister}
\bibfield{author}{\bibinfo{person}{R.W. Brandt}, \bibinfo{person}{E. Drenth},
  \bibinfo{person}{M. Montforts}, \bibinfo{person}{R.H.P. Proos},
  \bibinfo{person}{I.M. Roorda}, {and} \bibinfo{person}{R. Wiemer}.}
  \bibinfo{year}{1992}\natexlab{}.
\newblock \bibinfo{booktitle}{\emph{{Archeologisch Basisregister}}}.
\newblock \bibinfo{type}{{T}echnical {R}eport}.
  \bibinfo{institution}{Rijksdienst voor Cultureel Erfgoed},
  \bibinfo{address}{Amersfoort}.
\newblock


\bibitem[\protect\citeauthoryear{Byrne and Klein}{Byrne and Klein}{2010}]%
        {Byrne2010AutomaticText}
\bibfield{author}{\bibinfo{person}{Kate Byrne} {and} \bibinfo{person}{Ewan
  Klein}.} \bibinfo{year}{2010}\natexlab{}.
\newblock \showarticletitle{{Automatic Extraction of Archaeological Events from
  Text}}. In \bibinfo{booktitle}{\emph{Making History Interactive: Computer
  Applications and Quantitative Methods in Archaeology 2009}},
  \bibfield{editor}{\bibinfo{person}{B.~Frischer},
  \bibinfo{person}{J~Crawford}, \bibinfo{person}{}, {and}
  \bibinfo{person}{D~Koller}} (Eds.). \bibinfo{publisher}{BAR International
  Series 2079}, \bibinfo{address}{Oxford}, \bibinfo{pages}{48--56}.
\newblock


\bibitem[\protect\citeauthoryear{Capannini, Nardini, Perego, and
  Silvestri}{Capannini et~al\mbox{.}}{2011}]%
        {Capannini2011EfficientResults}
\bibfield{author}{\bibinfo{person}{Gabriele Capannini},
  \bibinfo{person}{Franco~Maria Nardini}, \bibinfo{person}{Raffaele Perego},
  {and} \bibinfo{person}{Fabrizio Silvestri}.} \bibinfo{year}{2011}\natexlab{}.
\newblock \showarticletitle{{Efficient diversification of web search results}}.
  In \bibinfo{booktitle}{\emph{Proceedings of the VLDB Endowment}}.
  \bibinfo{address}{Seattle, Washington}, \bibinfo{pages}{451 -- 459}.
\newblock
\showISSN{21508097}
\urldef\tempurl%
\url{https://doi.org/10.14778/1988776.1988781}
\showDOI{\tempurl}


\bibitem[\protect\citeauthoryear{Carpineto and Romano}{Carpineto and
  Romano}{2012}]%
        {Carpineto2012ARetrieval}
\bibfield{author}{\bibinfo{person}{Claudio Carpineto} {and}
  \bibinfo{person}{Giovanni Romano}.} \bibinfo{year}{2012}\natexlab{}.
\newblock \showarticletitle{{A survey of automatic query expansion in
  information retrieval}}.
\newblock \bibinfo{journal}{\emph{Comput. Surveys}} \bibinfo{volume}{44},
  \bibinfo{number}{1} (\bibinfo{date}{1} \bibinfo{year}{2012}).
\newblock
\showISSN{03600300}
\urldef\tempurl%
\url{https://doi.org/10.1145/2071389.2071390}
\showDOI{\tempurl}


\bibitem[\protect\citeauthoryear{Chan, Schweter, and M{\"{o}}ller}{Chan
  et~al\mbox{.}}{2021}]%
        {Chan2021GermansModel}
\bibfield{author}{\bibinfo{person}{Branden Chan}, \bibinfo{person}{Stefan
  Schweter}, {and} \bibinfo{person}{Timo M{\"{o}}ller}.}
  \bibinfo{year}{2021}\natexlab{}.
\newblock \showarticletitle{{German’s Next Language Model}}. In
  \bibinfo{booktitle}{\emph{Proceedings of the 28th International Conference on
  Computational Linguistics}}. \bibinfo{publisher}{International Committee on
  Computational Linguistics}, \bibinfo{address}{Barcelona, Spain},
  \bibinfo{pages}{6788--6796}.
\newblock
\urldef\tempurl%
\url{https://doi.org/10.18653/v1/2020.coling-main.598}
\showDOI{\tempurl}


\bibitem[\protect\citeauthoryear{Cheng, Bowden, Bhange, Goyal, Packer, and
  Javed}{Cheng et~al\mbox{.}}{2020}]%
        {Cheng2020AnSearch}
\bibfield{author}{\bibinfo{person}{Xiang Cheng}, \bibinfo{person}{Mitchell
  Bowden}, \bibinfo{person}{Bhushan~Ramesh Bhange}, \bibinfo{person}{Priyanka
  Goyal}, \bibinfo{person}{Thomas Packer}, {and} \bibinfo{person}{Faizan
  Javed}.} \bibinfo{year}{2020}\natexlab{}.
\newblock \showarticletitle{{An End-to-End Solution for Named Entity
  Recognition in eCommerce Search}}.
\newblock \bibinfo{journal}{\emph{arXiv}} (\bibinfo{date}{12}
  \bibinfo{year}{2020}).
\newblock
\urldef\tempurl%
\url{http://arxiv.org/abs/2012.07553}
\showURL{%
\tempurl}


\bibitem[\protect\citeauthoryear{Copara, Naderi, Knafou, Ruch, and
  Teodoro}{Copara et~al\mbox{.}}{2020}]%
        {Copara2020NamedModels}
\bibfield{author}{\bibinfo{person}{Jenny Copara}, \bibinfo{person}{Nona
  Naderi}, \bibinfo{person}{Julien Knafou}, \bibinfo{person}{Patrick Ruch},
  {and} \bibinfo{person}{Douglas Teodoro}.} \bibinfo{year}{2020}\natexlab{}.
\newblock \bibinfo{title}{{Named entity recognition in chemical patents using
  ensemble of contextual language models}}.
\newblock
\newblock
\showISSN{23318422}
\urldef\tempurl%
\url{http://arxiv.org/abs/2007.12569}
\showURL{%
\tempurl}


\bibitem[\protect\citeauthoryear{Corstius}{Corstius}{1981}]%
        {Corstius1981OpperlandseLetterkunde}
\bibfield{author}{\bibinfo{person}{Hugo~Brandt Corstius}.}
  \bibinfo{year}{1981}\natexlab{}.
\newblock \bibinfo{booktitle}{\emph{{Opperlandse taal-{\&} letterkunde}}}.
\newblock \bibinfo{publisher}{Querido}.
\newblock
\showISBNx{9789021451343}


\bibitem[\protect\citeauthoryear{Cowan, Zethelius, Luk, Baras, Ukarde, and
  Zhang}{Cowan et~al\mbox{.}}{2015}]%
        {Cowan2015NamedQueries}
\bibfield{author}{\bibinfo{person}{Brooke Cowan}, \bibinfo{person}{Sven
  Zethelius}, \bibinfo{person}{Brittany Luk}, \bibinfo{person}{Teodora Baras},
  \bibinfo{person}{Prachi Ukarde}, {and} \bibinfo{person}{Daodao Zhang}.}
  \bibinfo{year}{2015}\natexlab{}.
\newblock \showarticletitle{{Named entity recognition in travel-related search
  queries}}. In \bibinfo{booktitle}{\emph{Proceedings of the Twenty-Ninth AAAI
  Conference on Artificial Intelligence}}. \bibinfo{publisher}{AAAI Press},
  \bibinfo{address}{Austin, Texas}, \bibinfo{pages}{3935–3941}.
\newblock
\showISBNx{9781577357032}


\bibitem[\protect\citeauthoryear{de~Vries, van Cranenburgh, Bisazza, Caselli,
  van Noord, and Nissim}{de~Vries et~al\mbox{.}}{2019}]%
        {deVries2019BERTje:Model}
\bibfield{author}{\bibinfo{person}{Wietse de Vries}, \bibinfo{person}{Andreas
  van Cranenburgh}, \bibinfo{person}{Arianna Bisazza}, \bibinfo{person}{Tommaso
  Caselli}, \bibinfo{person}{Gertjan van Noord}, {and} \bibinfo{person}{Malvina
  Nissim}.} \bibinfo{year}{2019}\natexlab{}.
\newblock \bibinfo{title}{{BERTje: A Dutch BERT Model}}.
\newblock
\newblock
\urldef\tempurl%
\url{http://arxiv.org/abs/1912.09582}
\showURL{%
\tempurl}


\bibitem[\protect\citeauthoryear{Delobelle, Winters, and Berendt}{Delobelle
  et~al\mbox{.}}{2020}]%
        {Delobelle2020RobBERT:Model}
\bibfield{author}{\bibinfo{person}{Pieter Delobelle}, \bibinfo{person}{Thomas
  Winters}, {and} \bibinfo{person}{Bettina Berendt}.}
  \bibinfo{year}{2020}\natexlab{}.
\newblock \showarticletitle{{RobBERT: a Dutch RoBERTa-based Language Model}}.
  In \bibinfo{booktitle}{\emph{Findings of the Association for Computational
  Linguistics: EMNLP 2020}}. \bibinfo{publisher}{Association for Computational
  Linguistics}, \bibinfo{address}{Online}, \bibinfo{pages}{3255–3265}.
\newblock
\showISSN{23318422}
\urldef\tempurl%
\url{https://doi.org/10.18653/v1/2020.findings-emnlp.292}
\showDOI{\tempurl}


\bibitem[\protect\citeauthoryear{Devlin, Chang, Lee, and Toutanova}{Devlin
  et~al\mbox{.}}{2019}]%
        {Devlin2019BERT:Understanding}
\bibfield{author}{\bibinfo{person}{Jacob Devlin}, \bibinfo{person}{Ming~Wei
  Chang}, \bibinfo{person}{Kenton Lee}, {and} \bibinfo{person}{Kristina
  Toutanova}.} \bibinfo{year}{2019}\natexlab{}.
\newblock \showarticletitle{{BERT: Pre-training of deep bidirectional
  transformers for language understanding}}. In
  \bibinfo{booktitle}{\emph{Proceedings of the 2019 Conference of the North
  {\{}A{\}}merican Chapter of the Association for Computational Linguistics:
  Human Language Technologies, Volume 1 (Long and Short Papers)}},
  Vol.~\bibinfo{volume}{1}. \bibinfo{publisher}{Association for Computational
  Linguistics}, \bibinfo{address}{Minneapolis, Minnesota},
  \bibinfo{pages}{4171--4186}.
\newblock
\urldef\tempurl%
\url{https://doi.org/10.18653/v1/N19-1423}
\showDOI{\tempurl}


\bibitem[\protect\citeauthoryear{{ElasticSearch}}{{ElasticSearch}}{2018}]%
        {ElasticSearch2018TheoryScoring}
\bibfield{author}{\bibinfo{person}{{ElasticSearch}}.}
  \bibinfo{year}{2018}\natexlab{}.
\newblock \bibinfo{title}{{Theory Behind Relevance Scoring}}.
\newblock
\newblock
\urldef\tempurl%
\url{https://www.elastic.co/guide/en/elasticsearch/guide/current/scoring-theory.html}
\showURL{%
\tempurl}


\bibitem[\protect\citeauthoryear{Gibbs and Colley}{Gibbs and Colley}{2012}]%
        {Gibbs2012DigitalAustralia}
\bibfield{author}{\bibinfo{person}{Martin Gibbs} {and} \bibinfo{person}{Sarah
  Colley}.} \bibinfo{year}{2012}\natexlab{}.
\newblock \showarticletitle{{Digital Preservation,: Online access and
  historical archaeology 'grey literature' from New South Wales, Australia}}.
\newblock \bibinfo{journal}{\emph{Australian Archaeology}}
  \bibinfo{volume}{75} (\bibinfo{year}{2012}), \bibinfo{pages}{95--103}.
\newblock
\showISSN{03122417}
\urldef\tempurl%
\url{https://doi.org/10.1080/03122417.2012.11681957}
\showDOI{\tempurl}


\bibitem[\protect\citeauthoryear{Gormley and Tong}{Gormley and Tong}{2015}]%
        {Gormley2015Elasticsearch:Engine}
\bibfield{author}{\bibinfo{person}{C Gormley} {and} \bibinfo{person}{Z Tong}.}
  \bibinfo{year}{2015}\natexlab{}.
\newblock \bibinfo{booktitle}{\emph{{Elasticsearch: The Definitive Guide: A
  Distributed Real-Time Search and Analytics Engine}}}.
\newblock \bibinfo{publisher}{O'Reilly Media}, \bibinfo{address}{Sebastopol}.
\newblock


\bibitem[\protect\citeauthoryear{Guo, Xu, Cheng, and Li}{Guo
  et~al\mbox{.}}{2009}]%
        {Guo2009NamedQuery}
\bibfield{author}{\bibinfo{person}{Jiafeng Guo}, \bibinfo{person}{Gu Xu},
  \bibinfo{person}{Xueqi Cheng}, {and} \bibinfo{person}{Hang Li}.}
  \bibinfo{year}{2009}\natexlab{}.
\newblock \showarticletitle{{Named entity recognition in query}}. In
  \bibinfo{booktitle}{\emph{Proceedings - 32nd Annual International ACM SIGIR
  Conference on Research and Development in Information Retrieval, SIGIR
  2009}}. \bibinfo{publisher}{Association for Computing Machinery},
  \bibinfo{address}{Boston, Massachusetts}, \bibinfo{pages}{267--274}.
\newblock
\showISBNx{9781605584836}
\urldef\tempurl%
\url{https://doi.org/10.1145/1571941.1571989}
\showDOI{\tempurl}


\bibitem[\protect\citeauthoryear{Habermehl}{Habermehl}{2019}]%
        {Habermehl2019OverOnderzoek}
\bibfield{author}{\bibinfo{person}{Diederick Habermehl}.}
  \bibinfo{year}{2019}\natexlab{}.
\newblock \bibinfo{booktitle}{\emph{{Over zaaien en oogsten, de kwaliteit en
  bruikbaarheid van archeologische rapporten voor synthetiserend onderzoek}}}.
\newblock \bibinfo{type}{{T}echnical {R}eport}.
  \bibinfo{institution}{Rijksdienst voor Cultureel Erfgoed},
  \bibinfo{address}{Amersfoort}.
\newblock
\urldef\tempurl%
\url{https://www.cultureelerfgoed.nl/publicaties/publicaties/2019/01/01/over-zaaien-en-oogsten}
\showURL{%
\tempurl}


\bibitem[\protect\citeauthoryear{Hakala and Pyysalo}{Hakala and
  Pyysalo}{2019}]%
        {Hakala2019BiomedicalBERT}
\bibfield{author}{\bibinfo{person}{Kai Hakala} {and} \bibinfo{person}{Sampo
  Pyysalo}.} \bibinfo{year}{2019}\natexlab{}.
\newblock \showarticletitle{{Biomedical Named Entity Recognition with
  Multilingual BERT}}. In \bibinfo{booktitle}{\emph{Proceedings of The 5th
  Workshop on BioNLP Open Shared Tasks}}. \bibinfo{publisher}{Association for
  Computational Linguistics}, \bibinfo{address}{Hong Kong, China},
  \bibinfo{pages}{56–61}.
\newblock
\urldef\tempurl%
\url{https://doi.org/10.18653/v1/d19-5709}
\showDOI{\tempurl}


\bibitem[\protect\citeauthoryear{Jeffrey, Richards, Ciravegna, Waller, Chapman,
  and Zhang}{Jeffrey et~al\mbox{.}}{2009}]%
        {Jeffrey2009TheContext.}
\bibfield{author}{\bibinfo{person}{S Jeffrey}, \bibinfo{person}{J Richards},
  \bibinfo{person}{F Ciravegna}, \bibinfo{person}{S Waller}, \bibinfo{person}{S
  Chapman}, {and} \bibinfo{person}{Z Zhang}.} \bibinfo{year}{2009}\natexlab{}.
\newblock \showarticletitle{{The Archaeotools project: faceted classification
  and natural language processing in an archaeological context.}}
\newblock \bibinfo{journal}{\emph{Philosophical transactions. Series A,
  Mathematical, physical, and engineering sciences}} \bibinfo{volume}{367},
  \bibinfo{number}{1897} (\bibinfo{date}{6} \bibinfo{year}{2009}),
  \bibinfo{pages}{2507--19}.
\newblock
\urldef\tempurl%
\url{https://doi.org/10.1098/rsta.2009.0038}
\showDOI{\tempurl}


\bibitem[\protect\citeauthoryear{Kim and Lee}{Kim and Lee}{2020}]%
        {Kim2020KoreanBERT}
\bibfield{author}{\bibinfo{person}{Young~Min Kim} {and}
  \bibinfo{person}{Tae~Hoon Lee}.} \bibinfo{year}{2020}\natexlab{}.
\newblock \showarticletitle{{Korean clinical entity recognition from diagnosis
  text using BERT}}.
\newblock \bibinfo{journal}{\emph{BMC Medical Informatics and Decision Making}}
  \bibinfo{volume}{20}, \bibinfo{number}{S7} (\bibinfo{date}{9}
  \bibinfo{year}{2020}), \bibinfo{pages}{242}.
\newblock
\showISSN{14726947}
\urldef\tempurl%
\url{https://doi.org/10.1186/s12911-020-01241-8}
\showDOI{\tempurl}


\bibitem[\protect\citeauthoryear{Kudo and Richardson}{Kudo and
  Richardson}{2018}]%
        {Kudo2018SentencePiece:Processing}
\bibfield{author}{\bibinfo{person}{Taku Kudo} {and} \bibinfo{person}{John
  Richardson}.} \bibinfo{year}{2018}\natexlab{}.
\newblock \showarticletitle{{SentencePiece: A simple and language independent
  subword tokenizer and detokenizer for neural text processing}}. In
  \bibinfo{booktitle}{\emph{EMNLP 2018 - Conference on Empirical Methods in
  Natural Language Processing: System Demonstrations, Proceedings}}.
  \bibinfo{publisher}{Association for Computational Linguistics},
  \bibinfo{address}{Brussels, Belgium}, \bibinfo{pages}{66–71}.
\newblock
\showISBNx{9781948087858}
\urldef\tempurl%
\url{https://doi.org/10.18653/v1/d18-2012}
\showDOI{\tempurl}


\bibitem[\protect\citeauthoryear{Kuratov and Arkhipov}{Kuratov and
  Arkhipov}{2019}]%
        {Kuratov2019AdaptationLanguage}
\bibfield{author}{\bibinfo{person}{Yuri Kuratov} {and} \bibinfo{person}{Mikhail
  Arkhipov}.} \bibinfo{year}{2019}\natexlab{}.
\newblock \bibinfo{title}{{Adaptation of deep bidirectional multilingual
  transformers for Russian language}}.
\newblock
\newblock
\showISSN{23318422}
\urldef\tempurl%
\url{http://arxiv.org/abs/1905.07213}
\showURL{%
\tempurl}


\bibitem[\protect\citeauthoryear{Lafferty, Mccallum, Pereira, and
  Pereira}{Lafferty et~al\mbox{.}}{2001}]%
        {Lafferty2001ConditionalData}
\bibfield{author}{\bibinfo{person}{John Lafferty}, \bibinfo{person}{Andrew
  Mccallum}, \bibinfo{person}{Fernando C~N Pereira}, {and}
  \bibinfo{person}{Fernando Pereira}.} \bibinfo{year}{2001}\natexlab{}.
\newblock \showarticletitle{{Conditional Random Fields: Probabilistic Models
  for Segmenting and Labeling Sequence Data}}. In
  \bibinfo{booktitle}{\emph{Proc. 18th International Conf. on Machine
  Learning}}, \bibfield{editor}{\bibinfo{person}{Carla~E Brodley} {and}
  \bibinfo{person}{Danyluk~Andrea Pohoreckyj}} (Eds.).
  \bibinfo{publisher}{Morgan Kaufmann Publishers Inc.}, \bibinfo{address}{San
  Fransisco}, \bibinfo{pages}{282--289}.
\newblock


\bibitem[\protect\citeauthoryear{Lee, Yoon, Kim, Kim, Kim, So, and Kang}{Lee
  et~al\mbox{.}}{2019}]%
        {Lee2019BioBERT:Mining}
\bibfield{author}{\bibinfo{person}{Jinhyuk Lee}, \bibinfo{person}{Wonjin Yoon},
  \bibinfo{person}{Sungdong Kim}, \bibinfo{person}{Donghyeon Kim},
  \bibinfo{person}{Sunkyu Kim}, \bibinfo{person}{Chan~Ho So}, {and}
  \bibinfo{person}{Jaewoo Kang}.} \bibinfo{year}{2019}\natexlab{}.
\newblock \showarticletitle{{BioBERT: a pre-trained biomedical language
  representation model for biomedical text mining}}.
\newblock \bibinfo{journal}{\emph{Bioinformatics}} \bibinfo{volume}{36},
  \bibinfo{number}{4} (\bibinfo{date}{9} \bibinfo{year}{2019}),
  \bibinfo{pages}{1234--1240}.
\newblock
\showISSN{1367-4803}
\urldef\tempurl%
\url{https://doi.org/10.1093/bioinformatics/btz682}
\showDOI{\tempurl}


\bibitem[\protect\citeauthoryear{Li, Zhang, and Zhou}{Li et~al\mbox{.}}{2020}]%
        {Li2020ChineseMethods}
\bibfield{author}{\bibinfo{person}{Xiangyang Li}, \bibinfo{person}{Huan Zhang},
  {and} \bibinfo{person}{Xiao~Hua Zhou}.} \bibinfo{year}{2020}\natexlab{}.
\newblock \showarticletitle{{Chinese clinical named entity recognition with
  variant neural structures based on BERT methods}}.
\newblock \bibinfo{journal}{\emph{Journal of Biomedical Informatics}}
  \bibinfo{volume}{107} (\bibinfo{date}{7} \bibinfo{year}{2020}),
  \bibinfo{pages}{103422}.
\newblock
\showISSN{15320464}
\urldef\tempurl%
\url{https://doi.org/10.1016/j.jbi.2020.103422}
\showDOI{\tempurl}


\bibitem[\protect\citeauthoryear{Mikolov, Chen, Corrado, and Dean}{Mikolov
  et~al\mbox{.}}{2013}]%
        {Mikolov2013EfficientSpace}
\bibfield{author}{\bibinfo{person}{Tomas Mikolov}, \bibinfo{person}{Kai Chen},
  \bibinfo{person}{Greg Corrado}, {and} \bibinfo{person}{Jeffrey Dean}.}
  \bibinfo{year}{2013}\natexlab{}.
\newblock \showarticletitle{{Efficient estimation of word representations in
  vector space}}. In \bibinfo{booktitle}{\emph{1st International Conference on
  Learning Representations, ICLR 2013 - Workshop Track Proceedings}}.
\newblock


\bibitem[\protect\citeauthoryear{Moon, Awasthy, Ni, and Florian}{Moon
  et~al\mbox{.}}{2019}]%
        {Moon2019TowardsBERT}
\bibfield{author}{\bibinfo{person}{Taesun Moon}, \bibinfo{person}{Parul
  Awasthy}, \bibinfo{person}{Jian Ni}, {and} \bibinfo{person}{Radu Florian}.}
  \bibinfo{year}{2019}\natexlab{}.
\newblock \showarticletitle{{Towards Lingua Franca Named Entity Recognition
  with BERT}}.
\newblock \bibinfo{journal}{\emph{arXiv}} (\bibinfo{date}{11}
  \bibinfo{year}{2019}).
\newblock
\showISSN{23318422}
\urldef\tempurl%
\url{http://arxiv.org/abs/1912.01389}
\showURL{%
\tempurl}


\bibitem[\protect\citeauthoryear{Norvig}{Norvig}{2013}]%
        {Norvig2013EnglishRevisited}
\bibfield{author}{\bibinfo{person}{Peter Norvig}.}
  \bibinfo{year}{2013}\natexlab{}.
\newblock \bibinfo{title}{{English Letter Frequency Counts: Mayzner
  Revisited}}.
\newblock
\newblock
\urldef\tempurl%
\url{http://norvig.com/mayzner.html}
\showURL{%
\tempurl}


\bibitem[\protect\citeauthoryear{Nozza, Bianchi, and Hovy}{Nozza
  et~al\mbox{.}}{2020}]%
        {Nozza2020WhatModels}
\bibfield{author}{\bibinfo{person}{Debora Nozza}, \bibinfo{person}{Federico
  Bianchi}, {and} \bibinfo{person}{Dirk Hovy}.}
  \bibinfo{year}{2020}\natexlab{}.
\newblock \showarticletitle{{What the [MASK]? Making Sense of Language-Specific
  BERT Models}}.
\newblock \bibinfo{journal}{\emph{arXiv}} (\bibinfo{date}{3}
  \bibinfo{year}{2020}).
\newblock
\showISSN{23318422}
\urldef\tempurl%
\url{http://arxiv.org/abs/2003.02912}
\showURL{%
\tempurl}


\bibitem[\protect\citeauthoryear{Paijmans and Brandsen}{Paijmans and
  Brandsen}{2009}]%
        {Paijmans2009WhatDescriptions}
\bibfield{author}{\bibinfo{person}{Hans Paijmans} {and} \bibinfo{person}{Alex
  Brandsen}.} \bibinfo{year}{2009}\natexlab{}.
\newblock \showarticletitle{{What is in a Name: Recognizing Monument Names from
  Free-Text Monument Descriptions}}. In \bibinfo{booktitle}{\emph{Proceedings
  of the 18th Annual Belgian-Dutch Conference on Machine Learning
  (Benelearn)}}, \bibfield{editor}{\bibinfo{person}{M.G.J. van Erp},
  \bibinfo{person}{J.H. Stehouwer}, {and} \bibinfo{person}{M.~van Zaanen}}
  (Eds.). \bibinfo{publisher}{Tilburg centre for Creative Computing},
  \bibinfo{address}{Tilburg}, \bibinfo{pages}{2--6}.
\newblock


\bibitem[\protect\citeauthoryear{Paijmans and Brandsen}{Paijmans and
  Brandsen}{2010}]%
        {Paijmans2010}
\bibfield{author}{\bibinfo{person}{H. Paijmans} {and} \bibinfo{person}{A.
  Brandsen}.} \bibinfo{year}{2010}\natexlab{}.
\newblock \showarticletitle{{Searching in archaeological texts: Problems and
  solutions using an artificial intelligence approach}}.
\newblock \bibinfo{journal}{\emph{PalArch’s Journal Of Archaeology Of
  Egypt/Egyptology}} \bibinfo{volume}{7}, \bibinfo{number}{2}
  (\bibinfo{year}{2010}), \bibinfo{pages}{1--6}.
\newblock


\bibitem[\protect\citeauthoryear{Rabinowitz, Shaw, Buchanan, Golden, and
  Kansa}{Rabinowitz et~al\mbox{.}}{2016}]%
        {Rabinowitz2016MakingProject}
\bibfield{author}{\bibinfo{person}{Adam Rabinowitz}, \bibinfo{person}{Ryan
  Shaw}, \bibinfo{person}{Sarah Buchanan}, \bibinfo{person}{Patrick Golden},
  {and} \bibinfo{person}{Eric Kansa}.} \bibinfo{year}{2016}\natexlab{}.
\newblock \showarticletitle{{Making Sense of the Ways we make Sense of the
  Past: The PeriodO Project}}.
\newblock \bibinfo{journal}{\emph{Bulletin of the Institute of Classical
  Studies}} \bibinfo{volume}{59}, \bibinfo{number}{2} (\bibinfo{date}{12}
  \bibinfo{year}{2016}), \bibinfo{pages}{42--55}.
\newblock
\showISSN{0076-0730}
\urldef\tempurl%
\url{https://doi.org/10.1111/j.2041-5370.2016.12037.x}
\showDOI{\tempurl}


\bibitem[\protect\citeauthoryear{Rau}{Rau}{1991}]%
        {Rau1991ExtractingText}
\bibfield{author}{\bibinfo{person}{Lisa~F. Rau}.}
  \bibinfo{year}{1991}\natexlab{}.
\newblock \showarticletitle{{Extracting company names from text}}. In
  \bibinfo{booktitle}{\emph{Proceedings of the 7th IEEE Conference on
  Artificial Intelligence Applications}}. \bibinfo{publisher}{Publ by IEEE},
  \bibinfo{address}{Miami Beach, FL, USA}, \bibinfo{pages}{29--32}.
\newblock
\showISBNx{0818621354}
\urldef\tempurl%
\url{https://doi.org/10.1109/caia.1991.120841}
\showDOI{\tempurl}


\bibitem[\protect\citeauthoryear{{RCE}}{{RCE}}{2017}]%
        {RCE2017DeErfgoedmonitor}
\bibfield{author}{\bibinfo{person}{{RCE}}.} \bibinfo{year}{2017}\natexlab{}.
\newblock \bibinfo{title}{{De Erfgoedmonitor}}.
\newblock
\newblock
\urldef\tempurl%
\url{https://erfgoedmonitor.nl/indicatoren/archeologisch-onderzoek-aantal-onderzoeksmeldingen}
\showURL{%
\tempurl}


\bibitem[\protect\citeauthoryear{Richards, Tudhope, and Vlachidis}{Richards
  et~al\mbox{.}}{2015}]%
        {Richards2015TextReports}
\bibfield{author}{\bibinfo{person}{Julian Richards}, \bibinfo{person}{Douglas
  Tudhope}, {and} \bibinfo{person}{Andreas Vlachidis}.}
  \bibinfo{year}{2015}\natexlab{}.
\newblock \showarticletitle{{Text Mining in Archaeology: Extracting Information
  from Archaeological Reports}}.
\newblock In \bibinfo{booktitle}{\emph{Mathematics and Archaeology}},
  \bibfield{editor}{\bibinfo{person}{Juan~A. Barcelo} {and}
  \bibinfo{person}{Igor Bogdanovic}} (Eds.). \bibinfo{publisher}{CRC Press},
  \bibinfo{address}{Boca Raton}, \bibinfo{pages}{240--254}.
\newblock
\urldef\tempurl%
\url{https://doi.org/10.1201/b18530-15}
\showDOI{\tempurl}


\bibitem[\protect\citeauthoryear{Seok, Song, Park, Kim, and Kim}{Seok
  et~al\mbox{.}}{2016}]%
        {Seok2016NamedFeature}
\bibfield{author}{\bibinfo{person}{Miran Seok}, \bibinfo{person}{Hye~Jeong
  Song}, \bibinfo{person}{Chan~Young Park}, \bibinfo{person}{Jong~Dae Kim},
  {and} \bibinfo{person}{Yu~seop Kim}.} \bibinfo{year}{2016}\natexlab{}.
\newblock \showarticletitle{{Named entity recognition using word embedding as a
  feature}}.
\newblock \bibinfo{journal}{\emph{International Journal of Software Engineering
  and its Applications}}  \bibinfo{volume}{10} (\bibinfo{year}{2016}),
  \bibinfo{pages}{93 -- 104}.
\newblock
\showISSN{17389984}
\urldef\tempurl%
\url{https://doi.org/10.14257/ijseia.2016.10.2.08}
\showDOI{\tempurl}


\bibitem[\protect\citeauthoryear{Sien{\v{c}}nik}{Sien{\v{c}}nik}{2015}]%
        {sienvcnik2015adapting}
\bibfield{author}{\bibinfo{person}{Scharolta~Katharina Sien{\v{c}}nik}.}
  \bibinfo{year}{2015}\natexlab{}.
\newblock \showarticletitle{{Adapting word2vec to named entity recognition}}.
  In \bibinfo{booktitle}{\emph{Proceedings of the 20th Nordic Conference of
  Computational Linguistics, NODALIDA 2015}}.
  \bibinfo{publisher}{Link{\"{o}}ping University Electronic Press},
  \bibinfo{address}{Vilnius, Lithuania}, \bibinfo{pages}{239--243}.
\newblock


\bibitem[\protect\citeauthoryear{Song, Zhou, and He}{Song
  et~al\mbox{.}}{2011}]%
        {Song2011Post-rankingResults}
\bibfield{author}{\bibinfo{person}{Yang Song}, \bibinfo{person}{Dengyong Zhou},
  {and} \bibinfo{person}{Li~Wei He}.} \bibinfo{year}{2011}\natexlab{}.
\newblock \showarticletitle{{Post-ranking query suggestion by diversifying
  search results}}. In \bibinfo{booktitle}{\emph{SIGIR'11 - Proceedings of the
  34th International ACM SIGIR Conference on Research and Development in
  Information Retrieval}}. \bibinfo{publisher}{Association for Computing
  Machinery}, \bibinfo{address}{Beijing, China}, \bibinfo{pages}{815--824}.
\newblock
\showISBNx{9781450309349}
\urldef\tempurl%
\url{https://doi.org/10.1145/2009916.2010025}
\showDOI{\tempurl}


\bibitem[\protect\citeauthoryear{Soto, Olivas, and Prieto}{Soto
  et~al\mbox{.}}{2008}]%
        {Soto2008FuzzyRetrieval}
\bibfield{author}{\bibinfo{person}{Andrés Soto}, \bibinfo{person}{José~A.
  Olivas}, {and} \bibinfo{person}{Manuel~E. Prieto}.}
  \bibinfo{year}{2008}\natexlab{}.
\newblock \showarticletitle{{Fuzzy approach of synonymy and polysemy for
  information retrieval}}.
\newblock \bibinfo{journal}{\emph{Studies in Fuzziness and Soft Computing}}
  \bibinfo{volume}{224} (\bibinfo{year}{2008}), \bibinfo{pages}{179--198}.
\newblock
\showISBNx{9783540769729}
\showISSN{14349922}
\urldef\tempurl%
\url{https://doi.org/10.1007/978-3-540-76973-6_12}
\showDOI{\tempurl}


\bibitem[\protect\citeauthoryear{Souza, Nogueira, and Lotufo}{Souza
  et~al\mbox{.}}{2019}]%
        {Souza2019PortugueseBERT-CRF}
\bibfield{author}{\bibinfo{person}{Fábio Souza}, \bibinfo{person}{Rodrigo
  Nogueira}, {and} \bibinfo{person}{Roberto Lotufo}.}
  \bibinfo{year}{2019}\natexlab{}.
\newblock \showarticletitle{{Portuguese Named Entity Recognition using
  BERT-CRF}}.
\newblock \bibinfo{journal}{\emph{arXiv}} (\bibinfo{date}{9}
  \bibinfo{year}{2019}).
\newblock
\showISSN{23318422}
\urldef\tempurl%
\url{http://arxiv.org/abs/1909.10649}
\showURL{%
\tempurl}


\bibitem[\protect\citeauthoryear{Strubell, Ganesh, and McCallum}{Strubell
  et~al\mbox{.}}{2020}]%
        {Strubell2020EnergyNLP}
\bibfield{author}{\bibinfo{person}{Emma Strubell}, \bibinfo{person}{Ananya
  Ganesh}, {and} \bibinfo{person}{Andrew McCallum}.}
  \bibinfo{year}{2020}\natexlab{}.
\newblock \showarticletitle{{Energy and policy considerations for deep learning
  in NLP}}. In \bibinfo{booktitle}{\emph{ACL 2019 - 57th Annual Meeting of the
  Association for Computational Linguistics, Proceedings of the Conference}}.
  \bibinfo{publisher}{Association for Computational Linguistics},
  \bibinfo{address}{Florence, Italy}, \bibinfo{pages}{3645–3650}.
\newblock
\showISBNx{9781950737482}
\urldef\tempurl%
\url{https://doi.org/10.18653/v1/p19-1355}
\showDOI{\tempurl}


\bibitem[\protect\citeauthoryear{Talboom}{Talboom}{2017}]%
        {TalboomLeontien2017Itdo}
\bibfield{author}{\bibinfo{person}{Leontien Talboom}.}
  \bibinfo{year}{2017}\natexlab{}.
\newblock \emph{\bibinfo{title}{{Improving the discoverability of
  zooarchaeological data with the help of Natural Language Processing.}}}
\newblock \bibinfo{thesistype}{Ph.D. Dissertation}. \bibinfo{school}{University
  of York}.
\newblock


\bibitem[\protect\citeauthoryear{Tikhomirov, Loukachevitch, Sirotina, and
  Dobrov}{Tikhomirov et~al\mbox{.}}{2020}]%
        {Tikhomirov2020UsingDomain}
\bibfield{author}{\bibinfo{person}{Mikhail Tikhomirov}, \bibinfo{person}{N.
  Loukachevitch}, \bibinfo{person}{Anastasiia Sirotina}, {and}
  \bibinfo{person}{Boris Dobrov}.} \bibinfo{year}{2020}\natexlab{}.
\newblock \showarticletitle{{Using bert and augmentation in named entity
  recognition for cybersecurity domain}}. In \bibinfo{booktitle}{\emph{Natural
  Language Processing and Information Systems}}, Vol.~\bibinfo{volume}{12089
  LNCS}. \bibinfo{publisher}{Springer International Publishing},
  \bibinfo{address}{Cham}, \bibinfo{pages}{16--24}.
\newblock
\showISBNx{9783030513092}
\showISSN{16113349}
\urldef\tempurl%
\url{https://doi.org/10.1007/978-3-030-51310-8_2}
\showDOI{\tempurl}


\bibitem[\protect\citeauthoryear{Tunkelang}{Tunkelang}{2009}]%
        {Tunkelang2009FacetedSearch}
\bibfield{author}{\bibinfo{person}{Daniel Tunkelang}.}
  \bibinfo{year}{2009}\natexlab{}.
\newblock \showarticletitle{{Faceted Search}}.
\newblock \bibinfo{journal}{\emph{Synthesis Lectures on Information Concepts,
  Retrieval, and Services}} \bibinfo{volume}{1}, \bibinfo{number}{1}
  (\bibinfo{year}{2009}), \bibinfo{pages}{1--80}.
\newblock
\showISSN{1947-945X}
\urldef\tempurl%
\url{https://doi.org/10.2200/s00190ed1v01y200904icr005}
\showDOI{\tempurl}


\bibitem[\protect\citeauthoryear{Van~den Dries}{Van~den Dries}{2016}]%
        {VandenDries2016IsManagement}
\bibfield{author}{\bibinfo{person}{Monique Van~den Dries}.}
  \bibinfo{year}{2016}\natexlab{}.
\newblock \showarticletitle{{Is everybody happy? User satisfaction after ten
  years of quality management in European archaeological heritage management}}.
  In \bibinfo{booktitle}{\emph{When Valletta meets Faro, the reality of
  European archaeology in the 21st century, proceedings of the International
  Conference}}, \bibfield{editor}{\bibinfo{person}{P~Florjanowicz}} (Ed.).
  \bibinfo{publisher}{Archaeolingua}, \bibinfo{address}{Lisbon},
  \bibinfo{pages}{126--135}.
\newblock


\bibitem[\protect\citeauthoryear{Virtanen, Kanerva, Ilo, Luoma, Luotolahti,
  Salakoski, Ginter, and Pyysalo}{Virtanen et~al\mbox{.}}{2019}]%
        {Virtanen2019MultilingualFinnish}
\bibfield{author}{\bibinfo{person}{Antti Virtanen}, \bibinfo{person}{Jenna
  Kanerva}, \bibinfo{person}{Rami Ilo}, \bibinfo{person}{Jouni Luoma},
  \bibinfo{person}{Juhani Luotolahti}, \bibinfo{person}{Tapio Salakoski},
  \bibinfo{person}{Filip Ginter}, {and} \bibinfo{person}{Sampo Pyysalo}.}
  \bibinfo{year}{2019}\natexlab{}.
\newblock \showarticletitle{{Multilingual is not enough: BERT for Finnish}}.
\newblock \bibinfo{journal}{\emph{arXiv}} (\bibinfo{date}{12}
  \bibinfo{year}{2019}).
\newblock
\urldef\tempurl%
\url{http://arxiv.org/abs/1912.07076}
\showURL{%
\tempurl}


\bibitem[\protect\citeauthoryear{Vlachidis, Binding, May, and
  Tudhope}{Vlachidis et~al\mbox{.}}{2013}]%
        {Vlachidis2013AutomaticLiterature}
\bibfield{author}{\bibinfo{person}{Andreas Vlachidis}, \bibinfo{person}{Ceri
  Binding}, \bibinfo{person}{Keith May}, {and} \bibinfo{person}{Douglas
  Tudhope}.} \bibinfo{year}{2013}\natexlab{}.
\newblock \showarticletitle{{Automatic metadata generation in an archaeological
  digital library: Semantic annotation of grey literature}}.
\newblock \bibinfo{journal}{\emph{Studies in Computational Intelligence}}
  \bibinfo{volume}{458} (\bibinfo{year}{2013}), \bibinfo{pages}{187--202}.
\newblock
\showISBNx{9783642343988}
\showISSN{1860949X}
\urldef\tempurl%
\url{https://doi.org/10.1007/978-3-642-34399-5_10}
\showDOI{\tempurl}


\bibitem[\protect\citeauthoryear{Vlachidis, Tudhope, Wansleeben, Azzopardi,
  Green, Xia, and Wright}{Vlachidis et~al\mbox{.}}{2017}]%
        {Vlachidis2017}
\bibfield{author}{\bibinfo{person}{A Vlachidis}, \bibinfo{person}{D Tudhope},
  \bibinfo{person}{M Wansleeben}, \bibinfo{person}{J Azzopardi},
  \bibinfo{person}{K Green}, \bibinfo{person}{L Xia}, {and} \bibinfo{person}{H
  Wright}.} \bibinfo{year}{2017}\natexlab{}.
\newblock \bibinfo{booktitle}{\emph{{D16.4: Final Report on Natural Language
  Processing}}}.
\newblock \bibinfo{type}{{T}echnical {R}eport}. \bibinfo{institution}{ARIADNE}.
\newblock
\urldef\tempurl%
\url{http://legacy.ariadne-infrastructure.eu/wp-content/uploads/2019/01/D16.4_Final_Report_on_Natural_Language_Processing_Final.pdf}
\showURL{%
\tempurl}


\bibitem[\protect\citeauthoryear{Wen, Vasthimal, Lu, Wang, and Guo}{Wen
  et~al\mbox{.}}{2019}]%
        {Wen2019BuildingSearch}
\bibfield{author}{\bibinfo{person}{Musen Wen}, \bibinfo{person}{Deepak~Kumar
  Vasthimal}, \bibinfo{person}{Alan Lu}, \bibinfo{person}{Tian Wang}, {and}
  \bibinfo{person}{Aimin Guo}.} \bibinfo{year}{2019}\natexlab{}.
\newblock \showarticletitle{{Building large-scale deep learning system for
  entity recognition in e-commerce search}}. In \bibinfo{booktitle}{\emph{BDCAT
  2019 - Proceedings of the 6th IEEE/ACM International Conference on Big Data
  Computing, Applications and Technologies}}. \bibinfo{publisher}{Association
  for Computing Machinery, Inc}, \bibinfo{address}{New York, New York, USA},
  \bibinfo{pages}{149--154}.
\newblock
\showISBNx{9781450370165}
\urldef\tempurl%
\url{https://doi.org/10.1145/3365109.3368765}
\showDOI{\tempurl}


\bibitem[\protect\citeauthoryear{Wolf, Debut, Sanh, Chaumond, Delangue, Moi,
  Cistac, Rault, Louf, Funtowicz, Davison, Shleifer, von Platen, Ma, Jernite,
  Plu, Xu, Le~Scao, Gugger, Drame, Lhoest, and Rush}{Wolf
  et~al\mbox{.}}{2020}]%
        {Wolf2020Transformers:Processing}
\bibfield{author}{\bibinfo{person}{Thomas Wolf}, \bibinfo{person}{Lysandre
  Debut}, \bibinfo{person}{Victor Sanh}, \bibinfo{person}{Julien Chaumond},
  \bibinfo{person}{Clement Delangue}, \bibinfo{person}{Anthony Moi},
  \bibinfo{person}{Pierric Cistac}, \bibinfo{person}{Tim Rault},
  \bibinfo{person}{Remi Louf}, \bibinfo{person}{Morgan Funtowicz},
  \bibinfo{person}{Joe Davison}, \bibinfo{person}{Sam Shleifer},
  \bibinfo{person}{Patrick von Platen}, \bibinfo{person}{Clara Ma},
  \bibinfo{person}{Yacine Jernite}, \bibinfo{person}{Julien Plu},
  \bibinfo{person}{Canwen Xu}, \bibinfo{person}{Teven Le~Scao},
  \bibinfo{person}{Sylvain Gugger}, \bibinfo{person}{Mariama Drame},
  \bibinfo{person}{Quentin Lhoest}, {and} \bibinfo{person}{Alexander Rush}.}
  \bibinfo{year}{2020}\natexlab{}.
\newblock \showarticletitle{{Transformers: State-of-the-Art Natural Language
  Processing}}. In \bibinfo{booktitle}{\emph{Proceedings of the 2020 Conference
  on Empirical Methods in Natural Language Processing: System Demonstrations}}.
  \bibinfo{publisher}{Association for Computational Linguistics},
  \bibinfo{address}{Stroudsburg, PA, USA}, \bibinfo{pages}{38--45}.
\newblock
\urldef\tempurl%
\url{https://doi.org/10.18653/v1/2020.emnlp-demos.6}
\showDOI{\tempurl}


\bibitem[\protect\citeauthoryear{Wu and Dredze}{Wu and Dredze}{2020}]%
        {Wu2020AreBERT}
\bibfield{author}{\bibinfo{person}{Shijie Wu} {and} \bibinfo{person}{Mark
  Dredze}.} \bibinfo{year}{2020}\natexlab{}.
\newblock \showarticletitle{{Are All Languages Created Equal in Multilingual
  BERT?}}. In \bibinfo{booktitle}{\emph{Proceedings of the 5th Workshop on
  Representation Learning for NLP}}. \bibinfo{publisher}{Association for
  Computational Linguistics}, \bibinfo{address}{Stroudsburg, PA, USA},
  \bibinfo{pages}{120--130}.
\newblock
\urldef\tempurl%
\url{https://doi.org/10.18653/v1/2020.repl4nlp-1.16}
\showDOI{\tempurl}


\bibitem[\protect\citeauthoryear{Yamada, Asai, Shindo, Takeda, and
  Matsumoto}{Yamada et~al\mbox{.}}{2020}]%
        {Yamada2020LUKE:Self-attention}
\bibfield{author}{\bibinfo{person}{Ikuya Yamada}, \bibinfo{person}{Akari Asai},
  \bibinfo{person}{Hiroyuki Shindo}, \bibinfo{person}{Hideaki Takeda}, {and}
  \bibinfo{person}{Yuji Matsumoto}.} \bibinfo{year}{2020}\natexlab{}.
\newblock \showarticletitle{{LUKE: Deep Contextualized Entity Representations
  with Entity-aware Self-attention}}. In \bibinfo{booktitle}{\emph{Proceedings
  of the 2020 Conference on Empirical Methods in Natural Language Processing
  (EMNLP)}}. \bibinfo{publisher}{Association for Computational Linguistics},
  \bibinfo{address}{Stroudsburg, PA, USA}, \bibinfo{pages}{6442--6454}.
\newblock
\urldef\tempurl%
\url{https://doi.org/10.18653/v1/2020.emnlp-main.523}
\showDOI{\tempurl}


\end{thebibliography}


\end{document}